\documentclass[prd,eqsecnum,twocolumn,amsfonts,amssymb]{revtex4}

\usepackage{graphicx}

\usepackage{bm}

\newcommand{\beq}{\begin{equation}}
\newcommand{\eeq}{\end{equation}}
\newcommand{\beqs}{\begin{eqnarray}}
\newcommand{\eeqs}{\end{eqnarray}}
\newcommand{\lsim}{\mathrel{\raisebox{-
.6ex}{$\stackrel{\textstyle<}{\sim}$}}}

\newcommand{\drawsquare}[2]{\hbox{%
\rule{#2pt}{#1pt}\hskip-#2pt
\rule{#1pt}{#2pt}\hskip-#1pt
\rule[#1pt]{#1pt}{#2pt}}\rule[#1pt]{#2pt}{#2pt}\hskip-#2pt
\rule{#2pt}{#1pt}}
\newcommand{\fund}{\raisebox{-.5pt}{\drawsquare{6.5}{0.4}}}

\begin{document}

\title{Scheme-Independent Series Expansions at an 
Infrared Zero of the Beta Function in Asymptotically Free Gauge Theories} 

\author{Thomas A. Ryttov$^a$ and Robert Shrock$^b$}

\affiliation{(a) \ CP$^3$-Origins and Danish Institute for Advanced Study \\
Southern Denmark University, Campusvej 55, Odense, Denmark}

\affiliation{(b) \ C. N. Yang Institute for Theoretical Physics \\
Stony Brook University, Stony Brook, NY 11794, USA }

\begin{abstract}

  We consider an asymptotically free vectorial gauge theory, with gauge group
  $G$ and $N_f$ fermions in a representation $R$ of $G$, having an infrared
  (IR) zero in the beta function at $\alpha_{IR}$. We present general formulas
  for scheme-independent series expansions of quantities, evaluated at
  $\alpha_{IR}$, as powers of an $N_f$-dependent expansion parameter,
  $\Delta_f$.  First, we apply these to calculate the derivative
  $d\beta/d\alpha$ evaluated at $\alpha_{IR}$, denoted $\beta'_{IR}$, which is
  equivalent to the anomalous dimension of the ${\rm Tr}(F_{\mu\nu}F^{\mu\nu})$
  operator, to order $\Delta_f^4$ for general $G$ and $R$, and to order
  $\Delta_f^5$ for $G={\rm SU}(3)$ and fermions in the fundamental
  representation. Second, we calculate the scheme-independent expansions of the
  anomalous dimension of the flavor-nonsinglet and flavor-singlet bilinear
  fermion antisymmetric Dirac tensor operators up to order $\Delta_f^3$.  The
  results are compared with rigorous upper bounds on anomalous dimensions of
  operators in conformally invariant theories. Our other results include an
  analysis of the limit $N_c \to \infty$, $N_f \to \infty$ with $N_f/N_c$
  fixed, calculation and analysis of Pad\'e approximants, and comparison with
  conventional higher-loop calculations of $\beta'_{IR}$ and anomalous
  dimensions as power series in $\alpha$.

\end{abstract}


\maketitle


\section{Introduction}
\label{intro_section}

The evolution of an asymptotically free gauge theory from the ultraviolet (UV)
to the infrared is of fundamental importance. The evolution of the running
gauge coupling $g=g(\mu)$, as a function of the Euclidean momentum scale,
$\mu$, is described by the renormalization-group (RG) beta function \cite{rg},
$\beta_g = dg/dt$ or equivalently, 
\beq
\beta = \frac{d\alpha}{dt} = \frac{g}{2\pi} \, \beta_g \ , 
\label{betadef}
\eeq
where $\alpha(\mu) = g(\mu)^2/(4\pi)$ and $dt=d\ln \mu$ (the argument $\mu$
will often be suppressed in the notation).  Here we consider an asymptotically
free (AF) vectorial gauge theory with non-Abelian, Yang-Mills gauge group $G$
and $N_f$ copies (flavors) of fermions $\psi_j$, $j=1,...,N_f$ transforming
according to the representation $R$ of $G$.  We take the fermions to be
massless, since a massive fermion with mass $m_0$ would be integrated out of
the effective field theory at scales $\mu < m_0$ \cite{ac} and hence would not
affect the infrared limit $\mu \to 0$ that we study here.

In an asymptotically free theory with sufficiently large fermion content, the
beta function has an infrared zero at $\alpha_{IR}$ that controls the UV to IR
evolution.  Here we consider vectorial theories of this type.  As the scale
$\mu$ decreases from large values in the UV to small values in the IR,
$\alpha(\mu)$ approaches $\alpha_{IR}$ as $\mu \to 0$.  The properties of the
theory at this IR zero of the beta function are of considerable interest. If
this IR zero of the beta function occurs at sufficiently weak coupling so that
the gauge interaction does not produce any spontaneous chiral symmetry breaking
(S$\chi$SB), then it is an exact IR fixed point (IRFP) of the renormalization
group. The theory thus exhibits scale invariance with anomalous dimensions for
various (gauge-invariant) operators.  In this infrared limit, the theory is in
a chirally symmetric, deconfined, non-Abelian Coulomb phase (NACP).  If, on the
other hand, as $\mu$ decreases and $\alpha(\mu)$ increases toward
$\alpha_{IR}$, there is a scale $\mu = \Lambda$ at which $\alpha(\mu)$ exceeds
a critical value, denoted $\alpha_{cr}$, then the gauge interaction produces a
nonzero chiral condensate, with associated spontaneous chiral symmetry breaking
and dynamical mass generation for the fermions.  These fermions are thus
integrated out of the low-energy effective field theory that is operative for
$\mu < \Lambda$.  In this case, $\alpha_{IR}$ is only an approximate IRFP. We
define $N_{f,cr}$ to be the critical value of $N_f$ such that if 
$N_f > N_{f,cr}$, then the (asymptotically free) theory does not undergo 
spontaneous chiral symmetry breaking. At the two-loop ($2\ell$) level, 
$\alpha_{IR,2\ell}=-4\pi b_1/b_2$, where $b_\ell$ denotes the $\ell$-loop 
coefficient in the beta
function (see Eqs. (\ref{beta}) and (\ref{alfir_2loop}) below), and since $b_1$
\cite{b1} and $b_2$ \cite{b2} are independent of the scheme used for
regularization and renormalization of the theory \cite{gross75}, it follows
that $\alpha_{IR,2\ell}$ is also scheme-independent.

Physical quantities evaluated at an infrared fixed point of the renormalization
group at $\alpha=\alpha_{IR}$ are of basic interest.  Since these are
physical, their exact values must be scheme-independent.  In conventional
computations of these quantities, first, one expresses them as series
expansions in powers of $\alpha$, calculated to $n$-loop order; second, one
computes the IR zero of the beta function, denoted $\alpha_{IR,n}$, to the same
$n$-loop order; and third, one sets $\alpha=\alpha_{IR,n}$ in the series
expansion for the given quantity to obtain its value at the IR zero of the beta
function to this $n$-loop order.  However, these conventional series expansions
in powers of $\alpha$, calculated to a finite order, are scheme-dependent
beyond the leading one or two terms.  Specifically, the terms in the beta
function are scheme-dependent at loop order $\ell \ge 3$ and the terms in an
anomalous dimension are scheme-dependent at loop order $\ell \ge 2$.  Indeed,
as is well-known, the presence of scheme-dependence in higher-order
perturbative calculations is a general property in quantum field theory. 

Clearly, it would be very valuable to have a calculational framework in which
these physical quantities evaluated at $\alpha=\alpha_{IR}$ are expressed as a
series expansion such that at every order in this expansion the result is
scheme-independent.  A key point that was noted early on \cite{b1,b2,gw2} is
that $\alpha_{IR}$ becomes small as the number $N_f$ of fermions increases
toward the value $N_{f,b1z}$ (given below in Eq. (\ref{nfb1z})) at which the
one-loop term in the beta function, $b_1$, passes through zero.  At the
two-loop level, $\alpha_{IR} \propto \Delta_f$, where
\beq
\Delta_f = N_{f,b1z}-N_f \ . 
\label{deltaf}
\eeq
Indeed, in a theory with $G={\rm SU}(N_c)$ and fermions in the fundamental
representation, in the limit $N_c \to \infty$ and $N_f \to \infty$ with
$N_f/N_c$ fixed, $\alpha_{IR}$ can be made arbitrarily small.  Hence, one can
envision reliable perturbative calculations of series expansions for physical
quantities at this IRFP \cite{gw2,b2} and, in particular, series expansions of
these quantities in powers of $\Delta_f$ for reasonably small $\Delta_f$
\cite{bz}. Because $\Delta_f$ is obviously scheme-independent, it follows that
this perturbative series expansion in powers of $\Delta_f$ is
scheme-independent. Some early work on this was reported in
\cite{bz,grunberg92}.  Recently, in \cite{gtr}, a procedure for calculating the
coefficients of this scheme-independent expansion was given for the anomalous
dimension of the (gauge-invariant) fermion bilinear at the IR zero of the beta
function, and the coefficients in this expansion were calculated up to order
$\Delta_f^3$ in a vectorial asymptotically free gauge theory with gauge group
$G$ and $N_f$ fermions in a representation $R$.  This work also presented an
analogous calculation for a theory with ${\cal N}=1$ supersymmetry to order
$\Delta_f^2$. The results were then evaluated in the case of SU($N_c$) with
fermions in the fundamental ($F$) representation, $R=F$, with Young tableau
$\fund$. In \cite{flir}, for $G={\rm SU}(3)$ and $R=F$, we calculated the
$n$-loop value of the squared coupling, $\alpha_{IR,n\ell}$ and the resultant
value of $\gamma_{\bar\psi\psi}$ to five-loop order, and in \cite{gsi} we
calculated the scheme-independent expansion of $\gamma_{\bar\psi\psi}$ for the
representations $R$ is theory to order $\Delta_f^4$, using five-loop inputs,
and performed an extrapolation to infinite order in $\Delta_f$ to estimate the
exact value of $\gamma_{\bar\psi\psi}$ as a function of $N_f$.  The improvement
in the knowledge of the anomalous dimension $\gamma_{\bar\psi\psi}$ obtained
from the scheme-independent series expansions in \cite{gtr,gsi} is valuable not
only for general field-theoretic purposes, but also because theories with large
anomalous dimensions of fermion bilinears may be relevant for ultraviolet
completions of the Standard Model. Indeed, there has been considerable interest
in theories that might produce large $\gamma_{\bar\psi\psi} \sim O(1)$
associated with an IR zero of the beta function and resultant quasi-conformal
behavior \cite{wtc}. In \cite{gsi} we also compared our results with recent
lattice measurements of $\gamma_{\bar\psi\psi}$.

In this paper we report a number of new results on scheme-independent series
expansions in powers of $\Delta_f$.  As noted, we consider an asymptotically
free vectorial gauge theory with gauge group $G$ and $N_f$ fermions in the
representation $R$.  First, we present general formulas for coefficients in the
scheme-independent expansion in powers of $\Delta_f$ of an arbitrary
(gauge-invariant) physical quantity evaluated at $\alpha_{IR}$.  We calculate
the scheme-independent expansion of the derivative of the beta function,
$\beta' = d\beta/d\alpha$, evaluated at $\alpha_{IR}$, denoted $\beta'_{IR}$,
to order $\Delta_f^4$.  As a consequence of the trace anomaly relation, in a
theory with massless fermions, $\beta'_{IR}$ is equivalent to
$\gamma_{_{F^2,IR}}$, the anomalous dimension, evaluated at $\alpha_{IR}$, of
the operator ${\rm Tr}(F_{\mu\nu}F^{\mu\nu})$, where $F^a_{\mu\nu}$ is the
non-Abelian Yang Mills field-strength tensor.  For the special case where the
gauge group is SU(3) and the fermions are in the fundamental (triplet)
representation, we compute this expansion to order $\Delta_f^5$. This SU(3)
theory corresponds to quantum chromodynamics (QCD) with $N_f$ massless quarks.
For general $G$ and $R$, we calculate the scheme-independent expansion
coefficients to order $\Delta_f^3$ for the anomalous dimension, evaluated at
$\alpha_{IR}$, of the flavor-nonsinglet and flavor-singlet fermion bilinear
Dirac tensor operators.  Since the $\Delta_f$ expansion starts at the upper end
of the non-Abelian Coulomb phase (NACP) at $\Delta_f=0$, i.e., $N_f=N_{f,b1z}$,
and extends downward in $N_f$ with increasing $\Delta_f$, we focus mainly on
the infrared behavior in the NACP.  We show that our scheme-independent
calculations of the anomalous dimensions of ${\rm Tr}(F_{\mu\nu} F^{\mu\nu})$
and fermion bilinear operators in the non-Abelian Coulomb phase obey respective
rigorous upper bounds for conformally invariant theories.  As part of our
analysis, we compare results for various quantities calculated via the
scheme-independent expansion with results calculated via a conventional
higher-loop scheme-dependent expansion.  Further, for the case with $G={\rm
  SU}(N_c)$ and fermions in the fundamental representation, we discuss the
limit $N_c \to \infty$ and $N_f \to \infty$ with $N_f/N_c$ fixed and finite.
From ratios of scheme-independent expansion coefficients for $\beta'_{IR}$,
$\gamma_{\bar\psi\psi,IR}$ and the anomalous dimension of the fermion bilinear
antisymmetric Dirac tensor operator, we show, in agreement with, and extending
\cite{gtr}, that the scheme-independent $\Delta_f$ expansion should be
reasonably accurate in the non-Abelian Coulomb phase.  As with our earlier
work, the present study is motivated by the value of the new results for a
basic understanding of the renormalization-group evolution of asymptotically
free gauge theories, and also may be relevant to ultraviolet completions of the
Standard Model.

The paper is organized as follows. Some relevant background and methods are
discussed in Section \ref{methods_section}. In Section \ref{kn_an_section} we
present explicit formulas for the calculations of certain coefficients ($a_n$
and $k_n$ in Eqs. (\ref{air_delta}) and (\ref{zeroeq})) that are needed for the
rest of our work. General scheme-independent expansion formulas for anomalous 
dimensions of operators are given in Section \ref{general_expansion_section}. 
In this section we also discuss rigorous upper bounds on anomalous dimensions
in a conformally invariant theory and their application here.
We give our new results on scheme-independent calculations
of $\beta'_{IR}$ in Section \ref{betaprime_section}.
In Section \ref{kappa_section} we extend the
analysis of the scheme-independent expansion of the anomalous dimension for the
$m=0$ fermion bilinear previously studied in \cite{gtr} and \cite{gsi} with
several new results.  These include calculations for the limit $N_c \to
\infty$, $N_f \to \infty$ with $N_f/N_c$ fixed and analyses of Pad\'e
approximants, with comparison to scheme-dependent higher-loop conventional
calculations.  Section \ref{gammat_section} presents scheme-independent
calculations of the anomalous dimension for the fermion bilinear
(flavor-nonsinglet and flavor-singlet) antisymmetric rank-2 Dirac tensor
operator.  Our conclusions are given in Section \ref{conclusion_section} and
some auxiliary formulas are listed in Appendix \ref{bellhatappendix}.


\section{Background and Methods} 
\label{methods_section}

The beta function of this theory has the series expansion
\beq
\beta = -2\alpha \sum_{\ell=1}^\infty b_\ell \, a^\ell =
-2\alpha \sum_{\ell=1}^\infty \bar b_\ell \, \alpha^\ell \ ,
\label{beta}
\eeq
where
\beq
a= \frac{g^2}{16\pi^2} = \frac{\alpha}{4\pi} \ , 
\label{adef}
\eeq
$b_\ell$ is the $\ell$-loop coefficient,
$\bar b_\ell = b_\ell/(4\pi)^\ell$, and we extract a minus sign for
convenience, so $b_1 > 0$ for asymptotic freedom. For analysis of an IR zero
of $\beta$, it is convenient to extract the $\alpha^2$ factor that gives
rise to the UV zero at $\alpha=0$ and define a reduced ($r$) beta function
\beq
\beta_r = \frac{\beta}
{\Big (-\frac{\alpha^2 b_1}{2\pi}\Big )} =
1 + \frac{1}{b_1} \sum_{\ell=2}^\infty b_\ell \, a^{\ell-1} \ .
\label{beta_reduced}
\eeq
The $n$-loop ($n\ell$) beta function, denoted $\beta_{n\ell}$ and reduced beta
function, denoted $\beta_{r,n\ell}$ are obtained from the respective
Eqs. (\ref{beta}) and (\ref{beta_reduced}) by changing the upper limit on the
$\ell$-loop summation from $\infty$ to $n$.  As noted above, $b_1$ and $b_2$
are scheme-independent (SI), while the $b_\ell$ with $\ell \ge 3$ are
scheme-dependent (SD) \cite{gross75}.  For a general gauge group $G$ and
fermion representation $R$, the coefficients $b_1$ and $b_2$ were calculated in
\cite{b1} and \cite{b2}, and $b_3$ and $b_4$ were calculated in \cite{b3} and
\cite{b4} (and checked in \cite{b4p}) in the commonly used mass-independent 
$\overline{\rm MS}$
scheme \cite{msbar}.  Recently, for $G={\rm SU}(3)$ and $R=F$, 
$b_5$ was calculated in \cite{b5su3}.  For reference
and to show our normalizations explicitly, $b_1$ and $b_2$ are listed in
Appendix \ref{bellhatappendix}. As $N_f$ increases, $b_1$ decreases through
positive values and vanishes with sign reversal at $N_f=N_{f,b1z}$, where
\beq
N_{f,b1z} = \frac{11C_A}{4T_f} 
\label{nfb1z}
\eeq
(the subscript $b1z$ means ``$b_1$ zero''), where $C_A$ and $T_f$ are
group-theoretic invariants \cite{casimir,nfintegral}.  The asymptotic freedom
condition therefore implies the upper bound $N_f < N_{f,b1z}$.  We denote the
interval $0 \le N_f < N_{f,b1z}$ as $I_{AF}$.

For $N_f$ close to, but less than, $N_{f,b1z}$, $b_2 < 0$, so the
two-loop beta function has an IR zero, at the value
\beq
\alpha_{IR,2\ell}=-\frac{\bar b_1}{\bar b_2} = -\frac{4\pi b_1}{b_2} \ . 
\label{alfir_2loop}
\eeq
In general, the $n$-loop beta function has a double UV zero at
$\alpha=0$ and $n-1$ zeros away from the origin.  Among the latter, the
smallest (real, positive) zero, if such a zero occurs, is the physical IR
zero, denoted $\alpha_{IR,n\ell}$.  As $N_f$ decreases from $N_{f,b1z}$, $b_2$
passes through zero to positive values as $N_f$ passes through the value
\beq
N_{f,b2z} = \frac{17C_A^2}{2T_f(5C_A+3C_f)} \ . 
\label{nfb2z}
\eeq
Hence, with $N_f$ formally extended
from nonnegative integers to nonnegative real numbers \cite{nfintegral}, 
$\beta_{2\ell}$ has an IR zero (IRZ) for $N_f$ in the interval
\beq
I_{IRZ}: \quad N_{f,b2z} < N_f < N_{f,b1z} \  .
\label{nfinterval}
\eeq
We denote this interval as $I_{IRZ}$. 

As $N_f$ decreases in this interval, $\alpha_{IR,2\ell}$ increases toward
strong coupling.  Hence, to study the IR zero for $N_f$ toward the middle and
lower part of $I_{IRZ}$ with reasonable accuracy, one requires higher-loop
calculations. These were performed in \cite{gkgg}-\cite{lnn} for
$\alpha_{IR,n\ell}$ and for the anomalous dimension of the fermion bilinear
operator.  Clearly, a perturbative calculation of the IR zero of
$\beta_{n\ell}$ is only reliable if the resultant $\alpha_{IR,n\ell}$ is not
excessively large. Moreover, since the $b_\ell$ with $\ell \ge 3$ are
scheme-dependent, it is also incumbent upon one to ascertain the degree of
sensitivity of the value obtained for $\alpha_{IR,n\ell}$ for $n \ge 3$ to the
scheme used for the calculation. This task was carried out in 
\cite{sch}-\cite{gracey2015}. One way to do this is to perform the calculation
of $\alpha_{IR,n\ell}$ in one scheme, say $\overline{\rm MS}$, apply a scheme
transformation to obtain the value of $\alpha_{IR,n\ell}$ in another scheme,
and compare how close the two values are.  As we discussed in 
\cite{sch}-\cite{sch2}, an acceptable scheme
transformation function must satisfy a set of conditions, and although these
are automatically satisfied in the local vicinity of the origin, $\alpha=0$ (as
in optimized schemes for perturbative QCD calculations), they are not
automatically satisfied, and indeed, are quite restrictive conditions, when one
applies the scheme transformation at an IR zero away from the origin.  
Anomalous dimensions of composite fermion operators for 
$G={\rm SU}(3)$ have been calculated in \cite{cftbaryons}.

The one-loop coefficient $b_1$ is a polynomial of degree 1 in $N_f$ and the
higher-loop coefficients $b_\ell$ with $\ell \ge 2$ are polynomials of degree
$\ell-1$ in $N_f$.  Let us define 
\beq
b_\ell^{(0)} = b_\ell\Big |_{N_f=N_{f,b1z}}
\label{bell0}
\eeq
and, for $r \ge 1$, 
\beq
b_\ell^{(r)} = \frac{d^r b_\ell}{(dN_f)^r}\Big |_{N_f=N_{f,b1z}}
             = (-1)^r \frac{d^r b_\ell}{(d\Delta_f)^r}\Big |_{\Delta_f=0} 
\label{bellr}
\eeq
Then one has the scheme-independent results 
\beq
b_1^{(0)}=0 
\label{b10}
\eeq
(which is equivalent to the definition of $N_{f,b1z}$), 
\beq
b_1^{(1)}=\frac{4T_f}{3} \ , 
\label{b11}
\eeq
\beq
b_2^{(0)} = -C_A(7C_A+11C_f) \equiv -C_A D \ , 
\label{b20}
\eeq
where
\beq
D = 7C_A+11C_f \ , 
\label{d}
\eeq
and
\beq
b_2^{(1)} = -\frac{4}{3}(5C_A+3C_f)T_f \ . 
\label{b21}
\eeq
It is convenient to introduce the definition (\ref{d}), since powers of 
$D$ occur in the denominators of the scheme-independent 
expansion coefficients of anomalous
dimensions of bilinear fermion Dirac tensor operators and of $d\beta/d\alpha$
evaluated at the IR zero of the beta function. 

Thus, one has the finite Taylor series expansions 
\beq
b_1 = b_1^{(1)}(N_f-N_{f,b1z}) = -b_1^{(1)}\Delta_f 
\label{b1delta}
\eeq
and, for $\ell \ge 2$, 
\beqs
& & b_\ell = \sum_{r=0}^{\ell-1} \frac{1}{r!} b_\ell^{(r)} (N_f-N_{f,b1z})^r
= \sum_{r=0}^{\ell-1} \frac{(-1)^r}{r!} b_\ell^{(r)} \Delta_f^r \ . \cr\cr
& & 
\label{belldelta_ellge2}
\eeqs
We write Eqs. (\ref{b1delta}) and (\ref{belldelta_ellge2}) in a unified 
manner as
\beq
b_\ell = \sum_{r=0}^{r_{\rm max}(\ell)}
\frac{(-1)^r}{r!} b_\ell^{(r)} \Delta_f^r \ , 
\label{belldelta}
\eeq
where $r_{\rm max}(1)=1$ and $r_{\rm max}(\ell)=\ell-1$ if $\ell \ge 2$.

It will also be useful to recall some basic properties of the theory regarding
global flavor symmetries.  Because the $N_f$ fermions are massless, the
Lagrangian is invariant under the classical global chiral flavor ($fl$)
symmetry $G_{fl,cl} = {\rm U}(N_f)_L \otimes {\rm U}(N_f)_R$, or equivalently, 
\beqs
&& G_{fl,cl} = {\rm SU}(N_f)_L \otimes {\rm SU}(N_f)_R \otimes 
{\rm U}(1)_V \otimes {\rm U}(1)_A \cr\cr
& & 
\label{gfl_classical}
\eeqs
(where $V$ and $A$ denote vector and axial-vector). The U(1)$_V$ represents
fermion number, which is conserved by the bilinear operators that we consider.
The ${\rm U}(1)_A$ symmetry is broken by instantons, so
the actual nonanomalous global flavor symmetry is 
\beq
G_{fl} = {\rm SU}(N_f)_L \otimes {\rm SU}(N_f)_R \otimes 
{\rm U}(1)_V \ . 
\label{gfl}
\eeq
This $G_{fl}$ symmetry is respected in the (deconfined) non-Abelian Coulomb
phase, since there is no spontaneous chiral symmetry in this phase.  As noted
before, we focus on this phase in the present work, since the
(scheme-independent) $\Delta_f$ expansion starts from the upper end of the
interval $I_{IRZ}$ in this phase where $\alpha_{IR} \to 0$ as
$\Delta_f \to 0$. In contrast, in the phase with confinement and spontaneous
chiral symmetry breaking, the gauge interaction produces a bilinear fermion 
condensate, which can be written as 
$\sum_{j=1}^{N_f} \bar\psi_j \psi_j$, and this breaks 
$G_{fl}$ to ${\rm SU}(N_f)_V \otimes {\rm U}(1)_V$.


\section{Calculation of the Series Expansion Coefficients $k_n$ and $a_n$}
\label{kn_an_section}

We know that the exact $\alpha_{IR}$ (and also the $n$-loop
approximation to this exact $\alpha_{IR}$) vanishes (linearly) as a
function of $\Delta_f$ and that it is analytic at $\Delta_f=0$, so we can 
expand it, or equivalently, $a_{IR}=\alpha_{IR}/(4\pi)$, as a series expansion
in this variable, $\Delta_f$.  We write 
\beq
a_{IR} = \sum_{j=1}^\infty a_j \Delta_f^j \ . 
\label{air_delta}
\eeq
(Note that $a_j$ as defined here is equal to $a_j/2$ in terms of the $a_j$ in
Eq. (8) of \cite{gtr}.)  

One calculates the coefficients $a_j$ in two steps.  First, one 
evaluates $\beta_r$ in
Eq. (\ref{beta_reduced}) at $\alpha=\alpha_{IR}$, where it vanishes. Since the
prefactor $-8\pi a_{IR}^2$ in Eq. (\ref{beta}) 
is nonzero in general (although it does vanish at
$\Delta_f=0$), it follows that
\beq
\sum_{\ell=1}^\infty b_\ell \, (a_{IR})^{\ell-1} = 0 \ . 
\label{betasumzero}
\eeq
One then substitutes the finite Taylor series expansions for $b_\ell$, 
and $a_{IR}$, Eqs. (\ref{b1delta}), (\ref{belldelta}), and (\ref{air_delta}), 
in Eq. (\ref{betasumzero}) and thereby obtains the equation
\beqs
\beta_r |_{\alpha=\alpha_{IR}} & = & 0 = \sum_{\ell=1}^\infty
\bigg [ \bigg (\sum_{r=0}^{r_{\rm max}(\ell)} 
b_\ell^{(r)} \Delta_f^r \bigg ) \, 
\bigg (\sum_{j=1}^\infty a_j \Delta_f^j \bigg )^\ell \ \bigg ] \cr\cr
& = & \sum_{n=1}^\infty k_n \Delta_f^n \ .  
\label{zeroeq}
\eeqs
The results for the first three $k_n$ were given in \cite{gtr} and are:
\beq
k_1 = a_1 b_2^{(0)} - b_1^{(1)} \ , 
\label{k1}
\eeq
\beq
k_2 = a_2 b_2^{(0)} + a_1^2 b_3^{(0)} - a_1 b_2^{(1)} \ , 
\label{k2}
\eeq
and
\beqs
& & k_3 = a_3 b_2^{(0)} + 2a_1a_2b_3^{(0)} + a_1^3 b_4^{(0)} 
- a_2 b_2^{(1)} - a_1^2 b_3^{(1)} \ . \cr\cr
& & 
\label{k3}
\eeqs
From Eq. (\ref{zeroeq}), it follows that the coefficient $a_n$ occurs linearly
in the expression for $k_n$, in the single term $a_n b_2^{(0)}$ \ \cite{gtr}. 
To further show the structural forms of the $k_n$, we give $k_4$ and $k_5$
here: 
\begin{widetext}
\beq
k_4=a_4b_2^{(0)}+(a_2^2+2a_1a_3)b_3^{(0)}+3a_1^2a_2 b_4^{(0)}+a_1^4b_5^{(0)} 
-a_3b_2^{(1)} - 2a_1a_2 b_3^{(1)} - a_1^3b_4^{(1)}+\frac{1}{2}a_1^2 b_3^{(2)} 
\label{k4}
\eeq
\beqs
k_5 & = & a_5b_2^{(0)}+2(a_1a_4+a_2a_3)b_3^{(0)}+3a_1(a_2^2+a_1a_3)b_4^{(0)} 
+ 4a_1^3a_2 b_5^{(0)}+a_1^5b_6^{(0)} \cr\cr
& - & a_4b_2^{(1)} - (a_2^2+2a_1a_3)b_3^{(1)}-3a_1^2a_2b_4^{(1)}
-a_1^4b_5^{(1)} + a_1a_2 b_3^{(2)} + \frac{1}{2}a_1^3b_4^{(2)} \ . 
\label{k5}
\eeqs
\end{widetext}
In addition to the property that $k_n$ contains a term $a_n b_2^{(0)}$, we 
remark on two other general properties of the $k_n$: (i) 
$k_n$ contains a term $a_1^n b_{n+1}^{(0)}$ (which coincides with the term
$a_nb_2^{(0)}$ if $n=1$) and (ii) if $n \ge 2$,
then $k_n$ contains a term $-a_{n-1}b_2^{(1)}$.

Next, one observes that in Eq. (\ref{zeroeq}), since $\Delta_f$ is variable, 
this implies that the coefficients $k_n$ of each
power $\Delta_f^n$ must vanish individually.  One can solve the equations
$k_n=0$ for the $a_n$. The solutions are unique because of the property that 
$a_n$ occurs linearly in $k_n$. The solutions for the $a_n$ with 
with $1 \le n \le 3$ were given in \cite{gtr}. Thus, the equation 
$k_1=0$ yields 
\beq
a_1 = \frac{b_1^{(1)}}{b_2^{(0)}} \ . 
\label{a1sol}
\eeq
One then substitutes this into the equation $k_2=0$ and solves for $a_2$,
obtaining 
\beq
a_2 = \frac{b_1^{(1)}}{(b_2^{(0)})^3}\,(b_2^{(0)} b_2^{(1)}-b_1^{(1)}b_3^{(0)})
\ . 
\label{a2}
\eeq
One then proceeds iteratively in the manner, substituting the solutions for the
$a_k$ with $1 \le k \le n-1$ in the equation $k_n=0$ and solving for $a_n$. For
$a_3$, this yields
\beqs
a_3 & = & \frac{b_1^{(1)}}{(b_2^{(0)})^5}\, 
\bigg [ (b_2^{(0)}b_2^{(1)})^2 -3 b_1^{(1)}b_2^{(0)}b_2^{(1)}b_3^{(0)} 
+ 2(b_1^{(1)}b_3^{(0)})^2 \cr\cr
&+& b_1^{(1)}(b_2^{(0)})^2 \, b_3^{(1)}-(b_1^{(1)})^2 \, b_2^{(0)} b_4^{(0)} 
\bigg ] \ . 
\label{a3}
\eeqs
In general, $a_n$ depends on the $b_\ell$ coefficients for 
$1 \le \ell \le n+1$. 
The $a_n$ with $1 \le n \le 3$ were all the coefficients of this type
that were needed in \cite{gtr} since the $b_\ell$ have only been computed for a
general gauge group $G$ and fermion representation $R$ up to $\ell=4$ loop
order.  These $a_n$ have a factorized structure with a prefactor
\beq
a_n \propto \frac{b_1^{(1)}}{(b_2^{(0)})^{2n-1}} \ . 
\label{an_prefactor}
\eeq
In \cite{gsi} we also calculated and presented the result for $a_4$ for the
specific case $G={\rm SU}(3)$ and fermion representation $R=F$, since we were
using the recent calculation of the five-loop coefficient $b_5$ for this case
in \cite{b5su3}. Here we give the general result for $a_4$ for arbitrary gauge
group $G$ and fermion representation $R$:
\begin{widetext}
\beqs
a_4 & = & \frac{b_1^{(1)}}{(b_2^{(0)})^7} \, 
\bigg [ (b_2^{(0)}b_2^{(1)})^3 - \frac{1}{2}b_1^{(1)}(b_2^{(0)})^4b_3^{(2)} 
+(b_1^{(1)})^2(b_2^{(0)})^3b_4^{(1)} 
-4( b_1^{(1)}b_2^{(0)})^2 \Big ( b_2^{(1)}b_4^{(0)}+b_3^{(0)} b_3^{(1)} \Big )
\cr\cr
&+& 3b_1^{(1)}(b_2^{(0)})^2b_2^{(1)} \Big ( b_2^{(0)}b_3^{(1)} 
-2b_2^{(1)}b_3^{(0)} \Big ) 
+10(b_1^{(1)})^2b_2^{(0)}b_2^{(1)}(b_3^{(0)})^2 
+ 5(b_1^{(1)})^3b_3^{(0)} \Big ( b_2^{(0)}b_4^{(0)}-(b_3^{(0)})^2 \Big ) 
\cr\cr
&-& (b_1^{(1)})^3(b_2^{(0)})^2b_5^{(0)} \bigg ] \ . 
\label{a4}
\eeqs
\end{widetext}
In the same manner, we have calculated $a_5$ by substituting our solutions 
for the $a_k$ with $1 \le k \le 4$ in Eq. (\ref{k5}), and so forth for higher
$a_k$.  


\section{Scheme-Independent Series Expansion for Anomalous Dimensions
  at $\alpha_{IR}$} 
\label{general_expansion_section}


Let us consider a (gauge-invariant) operator ${\cal O}$. 
Because of the interactions, the full scaling
dimension of this operator, denoted $D_{\cal O}$, differs from its free-field 
value, $D_{\cal O,{\rm free}}$:
\beq
D_{\cal O} = D_{\cal O,{\rm free}} - \gamma_{\cal O} \ , 
\label{Dgamma} 
\eeq
where $\gamma_{\cal O}$ is the anomalous dimension of the operator
\cite{gammaconvention}.  Since $\gamma_{\cal O}$ arises from the gauge
interaction, it can be expressed as a power series about $a=0$:
\beq
\gamma_{\cal O} = \sum_{\ell=1}^\infty c_{{\cal O},\ell} \, a^\ell \ , 
\label{gammaseries}
\eeq
where $c_{{\cal O},\ell}$ is the $\ell$-loop coefficient. 

The exact anomalous
dimension $\gamma_{{\cal O}}$ evaluated at a zero of the exact beta
function, denoted $\gamma_{{\cal O},IR}$, is a physical quantity and hence is 
scheme-independent.  This was shown formally for the fermion bilinear operator
${\cal O} = \bar\psi\psi$ in \cite{gross75}, and the proof given there can be
straightforwardly extended to other (gauge-invariant) operators ${\cal O}$.  
However, this scheme-independence is not
preserved in a finite-order perturbative calculation, owing to the
scheme-dependence of the $b_\ell$ for $\ell \ge 3$ and of the 
$c_{{\cal O},\ell}$ for $\ell \ge 2$.  

As mentioned above, a method for calculating $\gamma_{\bar\psi\psi,IR}$ 
as a perturbative series expansion in powers of $\Delta_f$ was presented in
\cite{gtr}, with the important property that at each
order of the expansion the resulting approximation to
$\gamma_{\bar\psi\psi,IR}$ is scheme-independent.  
We can calculate a scheme-independent series 
expansion in powers of $\Delta_f$ for the anomalous dimension 
$\gamma_{\cal O}$ of a general (gauge-invariant) operator ${\cal O}$, evaluated
at $\alpha_{IR}$ by taking the series (\ref{gammaseries}) and inserting the 
expansions of $c_{{\cal O},\ell}$ and $a_{IR}$ as functions of $\Delta_f$. 
An advantage of this type of
series expansion is that since $\Delta_f$ is scheme-independent, so is the
expansion for $\gamma_{\cal O}$, in contrast to the
expression of $\gamma_{\cal O}$ as a series in powers of 
$\alpha_{IR,n\ell}$.  

We proceed to give a generalization of the results of \cite{gtr} for the
anomalous dimension of an arbitrary (gauge-invariant) operator ${\cal O}$ in an
asymptotically free gauge theory with gauge group $G$ and $N_f$ fermions 
in the representation $R$, evaluated at $\alpha_{IR}$. 
We denote this anomalous dimension as $\gamma_{{\cal O},IR}$.  Specifically, 
we present a general method for calculating
a series expansion of $\gamma_{{\cal O},IR}$ in powers of $\Delta_f$.  

We begin with the series expansion (\ref{gammaseries}) and substitute the 
series expansions for the $c_{{\cal O},\ell}$ and for $a_{IR}$.  Let 
\beq
c_{{\cal O},\ell}^{(0)} = c_{{\cal O},\ell}\Big |_{N_f=N_{f,b1z}}
\label{cell0}
\eeq
and, for $r \ge 1$, 
\beq
c_{{\cal O},\ell}^{(r)} = 
         \frac{d^r c_{{\cal O},\ell}}{(dN_f)^r}\Big |_{N_f=N_{f,b1z}}
= (-1)^r \frac{d^r c_{{\cal O},\ell}}{(d\Delta_f)^r}\Big |_{\Delta_f=0} 
\label{cellr}
\eeq
Then 
\beqs
\gamma_{{\cal O},IR} & = & \sum_{\ell=1}^\infty
\bigg [ \bigg (\sum_r c_{{\cal O},\ell}^{(r)} \Delta_f^r \bigg ) \, 
\bigg (\sum_{j=1}^\infty a_j \Delta_f^j \bigg )^\ell \ \bigg ] \cr\cr
& = & \sum_{n=1}^\infty \kappa_{{\cal O},n} \Delta_f^n \ .  
\label{gammadeltagen}
\eeqs
We denote the value of $\gamma_{{\cal O},IR}$ obtained from this series 
calculated to order $O(\Delta_f^p)$, i.e., from the last line of 
Eq. (\ref{betaprime_delta}) with the upper
limit of the summand changed from $\infty$ to $p$, as 
$\gamma_{{\cal O},IR,\Delta_f^p}$. 

We calculate
\beq
\kappa_{{\cal O},1} = a_1 c_{{\cal O},1}^{(0)} \ , 
\label{kappa1gen}
\eeq
\beq
\kappa_{{\cal O},2} = a_2c_{{\cal O},1}^{(0)} + 
a_1^2 c_{{\cal O},2}^{(0)} \ , 
\label{kappa2gen}
\eeq
\beq
\kappa_{{\cal O},3} = a_3c_{{\cal O},1}^{(0)} 
+ 2a_1 a_2 c_{{\cal O},2}^{(0)} 
+ a_1^3    c_{{\cal O},3}^{(0)} 
+ a_1^2    c_{{\cal O},2}^{(1)} \ , 
\label{kappa3gen}
\eeq
\beqs
\kappa_{{\cal O},4} & = & 
            a_4 c_{{\cal O},1}^{(0)} 
+ (2a_1a_3+a_2^2)c_{{\cal O},2}^{(0)} 
+ 3a_1^2a_2      c_{{\cal O},3}^{(0)} \cr\cr
&+&a_1^4          c_{{\cal O},4}^{(0)}
+ 2a_1a_2        c_{{\cal O},2}^{(1)}
+ a_1^3          c_{{\cal O},3}^{(1)} \ , 
\label{kappa4gen}
\eeqs
etc. for $\kappa_{{\cal O},n}$ with $n \ge 5$.   To calculate 
$\kappa_{{\cal O},n}$, one needs to know the $a_j$ and $c_j$ for 
$1 \le j \le n$. These $\kappa_{{\cal O} ,n}$ have the following general
properties: (i) $\kappa_{{\cal O},n}$ contains the term 
$a_n c_{{\cal O},1}^{(0)}$ and (ii) $\kappa_{{\cal O},n}$ contains the term 
$a_1^n c_{{\cal O},n}^{(0)}$ (which coincides with (i) if $n=1$).

A relevant question concerns the range of applicability of the
scheme-independent series expansion (\ref{gammadeltagen}). We address this
question here.  As noted above, our analysis in this paper is focused on the
non-Abelian Coulomb phase, since there is no spontaneous symmetry breaking in
this phase, and hence a zero of the beta function is an exact IR fixed point of
the renormalization group. This means that the theory at this fixed point is 
scale-invariant.  A number of studies have concluded that in this case of an
exact IRFP in this asymptotically free gauge theory, scale invariance implies
the larger symmetry of conformal invariance \cite{scaleconformal,gammabound}.

We will use several methods to assess the range of
validity of the (scheme-independent) small-$\Delta_f$ expansion.  A general
comment is that the properties of the theory change qualitatively as $N_f$
decreases through the value $N_{f,cr}$ and spontaneous chiral symmetry breaking
occurs and the fermions gain dynamical masses.  In the (chirally symmetric)
non-Abelian Coulomb phase with $N_{f,cr} < N_f < N_{f,b1z}$ is clearly
qualitatively different from the confined phase with spontaneous chiral
symmetry breaking at smaller $N_f$ below $N_{f,cr}$.  Therefore, one does not,
in general, expect the small-$\Delta_f$ series expansion to hold below
$N_{f,cr}$.  Estimating the range of applicability of this expansion is thus
connected with estimating the value of $N_{f,cr}$. 

For this purpose, as in our previous work \cite{bvh,bfs,gtr}, we can apply
a rigorous upper bound on the anomalous dimension of an operator from the
unitarity of a conformal
field theory.  If the approximate calculation of the anomalous dimension of a
given quantity at a fixed value of $\Delta_f$, computed up to order
$\Delta_f^p$, yields a value that exceeds this upper bound, then we can infer
that the calculation is not applicable at this value of $\Delta_f$ or
equivalently, $N_f$.  In particular, this can give information on the extent of
the non-Abelian Coulomb phase and the value of $N_{f,cr}$.  This bound is
applicable whether or not the coefficients $\kappa_{{\cal O},n}$ are all of the
same sign, but it is most useful if these coefficients do have the same sign,
since in this case for a fixed $\Delta_f$ the anomalous dimension is a
monotonic function of the order to which the small-$\Delta_f$ series expansion
is calculated.

A second method that we shall use to estimate the range of applicability of the
series expansions in powers of $\Delta_f$ is the ratio test.  If a function
$f(z)$ has a Taylor series $f(z) = \sum_{n=1}^\infty s_n z^n$, then the ratio
test states that the series is (absolutely) convergent if $|z| < z_0$, where
\beq
z_0 = \lim_{n \to \infty} \frac{|s_n|}{|s_{n+1}|} \ . 
\label{z0}
\eeq
Our application of the ratio test here is only intended to give a rough
estimate of this range of applicability of the $\Delta_f$ series expansion
since (i) we do not assume that the $\Delta_f$ expansion is a Taylor series
expansion, and (ii) with only a few terms in the series for a given quantity,
we can compute only a few ratios of adjacent coefficients.

Finally, a third method that we shall use is to calculate 
$[p,q]$ Pad\'e approximants to the $\Delta_f$ series expansions. As rational
functions of $\Delta_f$, the approximants with $q \ge 1$ have poles, and the
nearest poles to the origin give one estimate of the range of validity of the
expansions.  


\subsection{Upper Bound on Anomalous Dimensions} 

We now state and apply the upper bound on the anomalous dimension of an
operator in a theory with scale invariance or conformal invariance. Recall that
a (finite-dimensional) representation of the Lorentz group is specified by the
set $(j_1,j_2)$, where $j_1$ and $j_2$ take on nonnegative integral or
half-integral values \cite{lorentzgroup}.  
A Lorentz scalar operator thus transforms as
$(0,0)$, a Lorentz vector as $(1/2,1/2)$, an antisymmetric tensor like the
field-strength tensor $F^a_{\mu\nu}$ as $(1,0) \oplus (0,1)$, etc.  Then the
requirement of unitarity in a scale-invariant theory (in four spacetime
dimensions) requires that the full dimension $D_{\cal O}$ of an operator (other
than the identity) must satisfy the lower bound \cite{gammabound}
\beq
D_{\cal O} \ge j_1 + j_2 + 1 \ . 
\label{dlowerbound}
\eeq
With the definition (\ref{Dgamma}), this is equivalent to the upper bound on
the anomalous dimension 
\beq
\gamma_{\cal O} \le D_{\cal O, {\rm free}} - (j_1+j_2+1) \ . 
\label{ogamma_upperbound}
\eeq

The case $(j_1,j_2)=(0,0)$ includes the Lorentz scalar operators 
$F^a_{\mu\nu}F^{a \ \mu\nu}$, and the flavor-nonsinglet and flavor-singlet 
fermion bilinear operators $\bar\psi T_b \psi$ and $\bar\psi\psi$, where here 
$T_b$ is an element of the Lie algebra of the global flavor symmetry group 
SU($N_f$).  Hence, first, since $D_{F^2,{\rm free}}=4$, it follows from 
(\ref{ogamma_upperbound}) that 
the anomalous dimension of the $F^a_{\mu\nu}F^{a \ \mu\nu}$, evaluated at
$\alpha_{IR}$, must satisfy 
\beq
\gamma_{_{F^2,IR}} \le 3 \ . 
\label{gammaff_upperbound}
\eeq
Second, (\ref{ogamma_upperbound}) implies that the (equal) anomalous dimensions
of the flavor-nonsinglet and flavor-singlet fermion bilinear operators
$\bar\psi T_b \psi$ and $\bar\psi\psi$ evaluated at $\alpha_{IR}$, denoted
$\gamma_{\bar\psi\psi,IR}$, must satisfy
\beq
\gamma_{\bar\psi\psi,IR} \le 2 \ . 
\label{gamma_upperbound}
\eeq
The flavor-nonsinglet and flavor-singlet fermion bilinear antisymmetric rank-2
Dirac tensor operators proportional to $\bar\psi T_b \sigma_{\mu\nu}\psi$ and 
$\bar\psi \sigma_{\mu\nu}\psi$ to be analyzed below correpond to the case
$(j_1,j_2)=(1,0) \oplus (0,1)$ (as is clear from the fact that they can 
couple to the non-Abelian field-strength tensor to form a Lorentz scalar).
Hence, with $j_1+j_2=1$ for $(j_1,j_2)=(1,0)$ or (0,1), the bound 
(\ref{ogamma_upperbound}) implies that the (equal) anomalous dimensions of 
these operators evaluated at $\alpha_{IR}$, denoted $\gamma_{_{T,IR}}$, must 
satisfy 
\beq
\gamma_{_{T,IR}} \le 1 \ . 
\label{gammat_upperbound}
\eeq

We have applied the upper bound (\ref{gamma_upperbound}) in our previous
calculations of $\gamma_{\bar\psi\psi,IR,n\ell}$ at the $n$-loop level, up to
$n=4$ loops \cite{bvh,bc,lnn,gtr,flir,gsi}.  We have also applied a
corresponding upper bound in \cite{bfs,lnn,gtr} 
on the anomalous dimension of the
(gauge-invariant) bilinear chiral superfield operator $\Phi \tilde \Phi$ in a
vectorial asymptotically free gauge theory with gauge group $G$, ${\cal N}=1$
supersymmetry, and $N_f$ pairs of chiral superfields $\Phi_j$ and
$\tilde \Phi_j$, $1 \le j \le N_f$, transforming according to the
representations $R$ and $\bar R$ of $G$ \cite{bfs,lnn}.  
A theory of particular interest is
the case $R=F$; here, $N_{f,b1z}=3N_c$ and the lower end of the conformal
phase is known, namely
$N_{f,cr}=(3/2)N_c$ \cite{nsvz,seiberg} (which is integral and hence physical
if $N_c$ is even).
This theory corresponds to supersymmetric QCD with massless
matter fields, and is often denoted SQCD.  In this case, the upper bound is
$\gamma_{\bar\psi\psi} \le 1$, and this is saturated at the lower end of the
non-Abelian Coulomb phase.  The scheme-independent expansion in \cite{gtr}
exhibited excellent agreement with this exact result.


\section{Scheme-Independent Calculation of $\beta'_{IR}$}
\label{betaprime_section}


\subsection{Calculation to Order $\Delta_f^4$ for General $G$ and $R$}
\label{betaprime_general_section}

An important property of an asymptotically free theory at an IR zero of 
the beta function (IRFP of the renormalization group) is is the derivative 
of this beta function evaluated at $\alpha=\alpha_{IR}$,
\beq
\beta'_{IR} = \frac{d\beta}{d\alpha}\Big |_{\alpha=\alpha_{IR}} \ . 
\label{betaprime}
\eeq
This is scheme-independent, as was proved in \cite{gross75}
\cite{higherderivs}.  In a theory with massless fermions, as considered here,
the trace of the energy-momentum tensor, $T^\mu_\mu$, satisfies the relation
\cite{traceanomaly}
\beq
T^\mu_\mu = \frac{\beta}{4 \alpha} \, F_{\mu\nu}^a F^{a \mu\nu} \ ,
\label{traceanomaly}
\eeq
where $F^a_{\mu\nu}=\partial_\mu A^a_\nu - \partial_\nu A_\mu^a +             
g c^{abc}A^b_\mu A^c_\nu$ 
is the gluon field strength tensor \cite{otherform}.
Since the energy-momentum tensor is conserved, its anomalous dimension is zero,
and its full dimension is equal to its free-field dimension, 4.  Consequently,
the full scaling dimension of the rescaled operator $F_{\mu\nu} F^{a \mu\nu}$, 
denoted $D_{F^2}$, satisfies
\beq
D_{F^2} = 4 + \beta' - \frac{2\beta}{\alpha} \ , 
\label{betaprimerelation}
\eeq
where we use the shorthand notation $F^2 \equiv F_{\mu\nu}^a F^{a \mu\nu}$
\cite{traceanomaly2,glueball}.  We denote the anomalous dimension of this
operator as $\gamma_{_{F^2}}$ and its evaluation at
$\alpha_{IR}$ as $\gamma_{_{F^2,IR}}$.  From Eq. (\ref{betaprimerelation}), it
follows that at a zero of the beta function away from the origin, in
particular, the IR zero of an asymptotically free gauge theory of interest here
at $\alpha=\alpha_{IR}$, the derivative $\beta'_{IR}$ is equivalent to the
anomalous dimension of the operator $F^a_{\mu\nu}F^{a \ \mu\nu}$
\cite{gammaconvention}:
\beq
\beta'_{IR} = -\gamma_{_{F^2,IR}} \ . 
\label{betaprime_fsquared_anomdim}
\eeq

From Eq. (\ref{beta}), one obtains the conventional series expansion for 
$\beta'_{IR}$ in powers of $\alpha$, or equivalently, $a$:
\beq
\beta'_{IR} = -2 \sum_{\ell=1}^\infty (\ell+1) \, b_\ell \, a_{IR}^\ell \ . 
\label{betaprime_series}
\eeq
We denote $\beta'_{IR,n\ell}$ as the $n$-loop truncation of this infinite 
series.  The two-loop value is scheme-independent \cite{bc}: 
\beqs
\beta'_{IR,2\ell} & = & -\frac{2b_1^2}{b_2} \cr\cr
& = & \frac{(11C_A-4T_fN_f)^2}{3[2(5C_A+3C_f)T_fN_f-17C_A^2]} \ ,
\label{betaprime_2loop}
\eeqs
which is positive for $N_f \in I_{IRZ}$. However, at the level of $n \ge 3$
loops, the quantity $\beta'_{IR,n\ell}$ is scheme-dependent.  This quantity was
calculated up to the four-loop level in \cite{bc,lnn}, using $b_3$ and $b_4$
computed in the $\overline{\rm MS}$ scheme from \cite{b3,b4} (for SU(3), see
also the four-loop study \cite{pallante}).

Here we calculate a scheme-independent expansion of $\beta'_{IR}$ in powers of
$\Delta_f$ to order $\Delta_f^4$ for general $G$ and $R$ and to the five-loop
level, i.e., order $\Delta_f^5$, for SU(3). For general $G$ and $R$, we
substitute the expansions of $b_\ell$ and $a_{IR}$, as series in $\Delta_f$, in
Eq. (\ref{betaprime_series}) to obtain
\beqs
\beta'_{IR} & = & -2 \sum_{\ell=1}^\infty (\ell+1) \, 
\bigg [ \bigg (\sum_{r=0}^{r_{\rm max}(\ell)} 
b_\ell^{(r)} \Delta_f^r \bigg ) \, 
\bigg (\sum_{j=1}^\infty a_j \Delta_f^j \bigg )^\ell \ \bigg ] \cr\cr
& = & \sum_{n=1}^\infty d_n \, \Delta_f^n \ . 
\label{betaprime_delta}
\eeqs
We denote the value of $\beta'_{IR}$ obtained from this series calculated to
order $\Delta_f^p$ as $\beta'_{IR,\Delta_f^p}$. The calculation of $d_n$
contains explicit dependence on the $b_\ell$ for $1 \le \ell \le n$ and on the
$a_j$ for $1 \le j \le n-1$; since $a_j$ depends on $b_\ell$ for $1 \le \ell
\le j+1$, it follows that the calculation of $d_n$ requires knowledge of
$b_\ell$ for $1 \le \ell \le n$.  Since the $b_\ell$ have been calculated for
general gauge group $G$ and fermion representation $R$ up to four-loop level,
we can thus calculate explicit expressions for the $d_n$ up to $n=4$.  For our
calculation, in addition to the scheme-independent results for $b_1$ and $b_2$
\cite{b1,b2}, we have used the expressions for $b_3$ and $b_4$ calculated in
the $\overline{\rm MS}$ scheme in \cite{b3,b4}. However, we stress that it does
not matter which scheme we use for $b_3$ and $b_4$, because the resulting
series expansion for $\beta'_{IR}$ in powers of $\Delta_f$ is
scheme-independent.

Substituting the $b_\ell^{(r)}$ and 
$a_j$ into these equations, we find the following results. First, 
\beq
d_1 = 0 \ , 
\label{d1}
\eeq
so that $\beta'_{IR}$ vanishes quadratically with $\Delta_f$ as
$\Delta_f \to 0$, i.e., as $N_f \to N_{f,b1z}$.  For $n \ge 2$, with 
the denominator factor $D=7C_A+11C_f$ as defined in Eq. (\ref{d}), 
we calculate 
\beq
d_2 = \frac{2^5 T_f^2}{3^2 C_A D} \ , 
\label{d2}
\eeq
\beq
d_3 = \frac{2^7 T_f^3(5C_A+3C_f)}{3^3 C_A^2 D^2} \ , 
\label{d3}
\eeq
and
\begin{widetext}
\beqs
d_4 &=& -\frac{2^3T_f^2}{3^6C_A^4 D^5} \, \bigg [
  C_A^5T_f^2(-412335+1241856\zeta_3) 
+ C_A^4T_f^2C_f(-310800+2661120\zeta_3) \cr\cr
&+& C_A^3T_f^2C_f^2(-217848-836352\zeta_3) + 
 C_A^3 \frac{d_R^{abcd}d_R^{abcd}}{d_A} \, (-2385152+5203968\zeta_3)
+C_A^2T_f^2C_f^3(-2855424-3066624\zeta_3) \cr\cr
&+&C_A^2T_f \, \frac{d_R^{abcd}d_A^{abcd}}{d_A} \, (630784-6150144\zeta_3) 
  +C_A^2C_f \, \frac{d_R^{abcd}d_R^{abcd}}{d_A} \, (-3748096+8177664\zeta_3)
\cr\cr
&+&191664C_AT_f^2C_f^4 
  +C_AT_f^2 \, \frac{d_A^{abcd}d_A^{abcd}}{d_A} \, (-35840+946176\zeta_3)
  +C_AT_fC_f \, \frac{d_R^{abcd}d_A^{abcd}}{d_A} \, (991232-9664512\zeta_3)
\cr\cr
&+&T_f^2C_f \frac{d_A^{abcd}d_A^{abcd}}{d_A} \, (-56320+1486848\zeta_3) \bigg ]
\ . 
\label{d4}
\eeqs
\end{widetext} 
Here, 
\beq
\zeta_s = \sum_{n=1}^\infty \frac{1}{n^s}
\label{zetas}
\eeq
is the Riemann zeta function, the quantities $C_A$, $C_f$, $T_f$ are 
group invariants, and the contractions 
$d_A^{abcd}d_A^{abcd}$, $d_R^{abcd}d_A^{abcd}$, 
$d_R^{abcd}d_R^{abcd}$ are additional group-theoretic quantities given in 
\cite{b4}, and $d_A$ is the dimension of the adjoint
representation of $G$. These calculations thus determine the quantity 
$\beta'_{IR}$ to order $\Delta_f^4$ for an arbitrary 
gauge group $G$ and fermion representation $R$. We have also calculated $d_5$,
but the expression is sufficiently lengthy that we do not include it here; 
however, we shall use it below. 

We note a general result on the signs of the first two nonzero coefficients in
the scheme-independent expansion for $\beta'_{IR}$: 
\beq
d_n > 0 \quad {\rm for} \ n=2, \ 3 \quad {\rm and \ arbitrary} \ G, \ R \ . 
\label{d12positive}
\eeq
These positivity results are clear from Eqs. (\ref{d2}) and (\ref{d3}). In 
contrast, there are terms of both signs in the large square bracket in the
expression for $d_4$, Eq. (\ref{d4}); 
for example, in the large square bracket in
Eq. (\ref{d4}), the coefficients of the $C_A^5T_f^2$ and $C_A^4T_f^2C_f$ 
terms are positive while the coefficient of the $C_A^3T_f^2C_f^2$ term is
negative, etc.  Indeed, we will show below in Eqs. (\ref{d4f}) and 
(\ref{d4adj}) that for $G={\rm SU}(N_c)$, $d_4$ is negative if $R=F$ and
positive if $R=adj$.  A summary of the sign results for these coefficients and
others is given in Table \ref{signs} for the case where $G={\rm SU}(N_c)$. 

In Table \ref{betaprime_values} we list the (scheme-independent) values that we
calculate for $\beta'_{IR,\Delta_f^p}$ with $2 \le p \le 4$ for 
the illustrative gauge groups $G={\rm SU}(2)$, SU(3), and SU(4), as 
functions of $N_f$ in the respective intervals $I_{IRZ}$ given in
Eq. (\ref{nfinterval}). For 
comparison, we list the $n$-loop values of 
$\beta'_{IR,n\ell}$ with the $2 \le n \le 4$, where 
$\beta'_{IR,3\ell}$ and $\beta'_{IR,4\ell}$ are computed in the 
$\overline{\rm MS}$ scheme. 
Values that exceed $\beta'_{IR} = 3$ are marked as
such. In the case of SU(3), we 
also include our calculation of $\beta'_{IR,\Delta_f^5}$. 


\subsection{Evaluation for $G={\rm SU}(N_c)$ and $R=F$ }
\label{betaprime_fundamental_section}

We proceed to evaluate our general formulas for the $d_n$ coefficients for a
case of particular interest, namely that in which the gauge group is 
$G={\rm SU}(N_c)$ with $N_f$ fermions in the fundamental representation, 
$R=F$. In
addition to Eq. (\ref{d1}), our general results (\ref{d2})-(\ref{d4}) yield
\beq
d_{2,{\rm SU}(N_c),F} = \frac{2^4}{3^2(25N_c^2-11)} \ , 
\label{d2f}
\eeq
\beq
d_{3,{\rm SU}(N_c),F} = \frac{2^5(13N_c^2-3)}{3^3N_c(25N_c^2-11)^2} \ , 
\label{d3f}
\eeq
and
\begin{widetext} 
\beqs
d_{4,{\rm SU}(N_c),F} & = & -\frac{2^4}{3^5N_c^2(25N_c^2-11)^5} \, \bigg [
N_c^8\Big (-366782+660000\zeta_3 \Big )+
N_c^6\Big (865400-765600\zeta_3 \Big ) \cr\cr
&+& N_c^4\Big (-1599316+2241888\zeta_3 \Big ) 
 +  N_c^2\Big (571516-894432\zeta_3 \Big ) + 3993 \ \bigg ] \ . 
\label{d4f}
\eeqs
\end{widetext}
As is evident, the coefficients $d_{2,{\rm SU}(N_c),F}$ and 
$d_{3,{\rm SU}(N_c),F}$ are positive-definite for all physical values of
$N_c$.  We find that $d_{4,{\rm SU}(N_c),F}$ is negative-definite for all
physical values of $N_c \ge 2$. 


\subsection{Calculation to $O(\Delta_f^5)$ for $G={\rm SU}(3)$ and $R=F$ }
\label{betaprime_su3_section}

\begin{figure}
  \begin{center}
    \includegraphics[height=6cm]{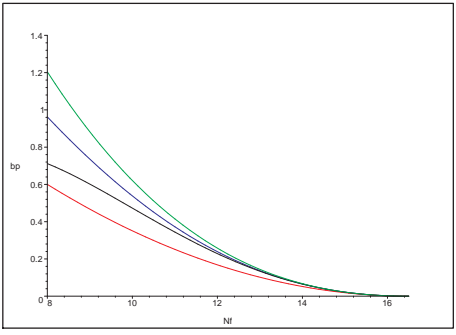}
  \end{center}
\caption{Plot of $\beta'_{IR,\Delta_f^p}$ for $2 \le p \le 5$ as a
  function of $N_f$ for the SU(3) theory with $N_f$ fermions in the fundamental
  representation. From bottom to top, the curves (with colors online)
 refer to  $\beta'_{IR,\Delta_f^2}$ (red), 
           $\beta'_{IR,\Delta_f^5}$ (black)
           $\beta'_{IR,\Delta_f^4}$ (blue),
           $\beta'_{IR,\Delta_f^3}$ (green).}
\label{betaprime_Delta_plot}
\end{figure}

For the special case where the gauge group is $G={\rm SU}(3)$ and the $N_f$
fermions are in the fundamental representation, $R=F$, we can make use of
the recent calculation of $b_5$ in the $\overline{\rm MS}$ scheme in
\cite{b5su3} to carry out the scheme-independent calculation of $\beta'_{IR}$
to one order higher than for general $G$ and $R$, namely to order $\Delta_f^5$.
We first give the special cases of our results in Eqs. (\ref{d1})-(\ref{d4})
for this theory. In addition to $d_{1,{\rm SU}(3),F}=0$, we find
\beq
d_{2,{\rm SU}(3),F}=\frac{8}{3^2 \cdot 107} =
0.830737 \times 10^{-2} \ , 
\label{d2su3f}
\eeq
\bigskip
\beqs
& & d_{3,{\rm SU}(3),F} = \frac{304}{3^3 \cdot (107)^2} 
                = 0.983427 \times 10^{-3} \ , \cr\cr
& & 
\label{d3su3f}
\eeqs
and
\beqs
d_{4,{\rm SU}(3),F}&=& \frac{633325687}{2 \cdot 3^6 \cdot (107)^5} 
- \frac{682880}{3^4 \cdot (107)^4} \, \zeta_3 \cr\cr
& = & -(0.463417 \times 10^{-4}) \ . 
\label{d4su3f}
\eeqs
For $d_{5,{\rm SU}(3),F}$ we calculate 
\begin{widetext}
\beqs
d_{5,{\rm SU}(3),F} &=& -\frac{66670528901419}{2 \cdot 3^9 \cdot (107)^7}
-   \frac{122882810048}{3^8 \cdot (107)^6} \, \zeta_3
+   \frac{196275200}{3^6 \cdot (107)^5} \, \zeta_5 \cr\cr
& = & - (0.564349 \times 10^{-5}) \ .
\label{d5su3}
\eeqs
In these equations we have indicated the simple factorizations of the
denominators that were already evident in the general analytic expressions
(\ref{d1})-(\ref{d4}). The numerators do not, in general, have such simple
systematic factorizations; for example, in $d_{4,{\rm SU}(3),F}$, the number
$633325687 = 227 \cdot 311 \cdot 8971$, etc.  We will also use this
factorization format, indicating the factorizations of the denominators, in
later equations.  Substituting these coefficients into
Eq. (\ref{betaprime_delta}), we have, to $O(\Delta_f^5)$,
\beqs
\beta'_{IR} & = & \Delta_f^2 \, \Big [ (0.830737 \times 10^{-2}) 
+ (0.983427 \times 10^{-3})\Delta_f 
- (0.463417 \times 10^{-4})\Delta_f^2 
- (0.564349 \times 10^{-5})\Delta_f^3 \Big ] \ , 
\label{betaprime_su3_quintic}
\eeqs
to the indicated floating-point accuracy. 
\end{widetext}
In Fig. \ref{betaprime_Delta_plot} we plot the values of $\beta'_{IR}$,
calculated to order $\Delta_f^p$ with $2 \le p \le 5$.  In the general
calculations of $\gamma_{\bar\psi\psi,IR}$ as a series in powers of $\Delta_f$
to maximal power $p=3$ (i.e., order $\Delta_f^3$) in \cite{gtr} and, for
$G={\rm SU}(3)$ and $R=F$, to maximal power $p=4$ in \cite{gsi}, it was found
that, for a fixed value of $N_f$, or equivalently, $\Delta_f$, in the interval
$I_{IRZ}$, these anomalous dimensions increased monotonically as a function of
$p$. This feature motivated our extrapolation to $p=\infty$ in \cite{gtr} to
obtain estimates for the exact $\gamma_{\bar\psi\psi,IR}$.  In contrast, here
we find that, for a fixed value of $N_f$, or equivalently, $\Delta_f$, in
$I_{IRZ}$, as a consequence of the fact that different coefficients $d_n$ do
not all have the same sign, $\beta'_{IR,\Delta_f^p}$ is not a monotonic
function of $p$. Because of this non-monotonicity, we do not attempt to
extrapolate our series to $p=\infty$.  Lattice measurements of
$\gamma_{_{F^2,IR}}$ or $\beta'_{IR}$ would be useful here (see also
\cite{pallante}). In particular, for $G={\rm SU}(3)$ and fermions in the
fundamental representation, the lattice measurements of $\gamma_{_{F^2,IR}}$
could be compared with our scheme-independent calculation of
$\beta'_{IR}$ to order $\Delta_f^5$, similar to the
comparison of our scheme-independent calculation of $\gamma_{\bar\psi\psi,IR}$
to order $\Delta_f^4$ (which also used five-loop inputs \cite{looplevel}) with
lattice results that we carried out in \cite{gsi}.

To get a rough estimate of the range of accuracy and applicability of the
series expansion for $\beta'_{IR}$, we can compute ratios of coefficients, as
discussed in connection with Eq. (\ref{z0}).  Thus, we have 
\beq
\frac{d_{2,{\rm SU}(3),F}}{d_{3,{\rm SU}(3),F}} = 8.447
\label{d2_over_d3}
\eeq
\beq
\frac{d_{3,{\rm SU}(3),F}}{|d_{4,{\rm SU}(3),F}|} = 21.221 , 
\label{d3_over_d4}
\eeq
and
\beq
\frac{|d_{4,{\rm SU}(3),F}|}{|d_{5,{\rm SU}(3),F}|} = 8.2115 
\label{d4_over_d5}
\eeq
Since $N_{f,b1z}=16.5$ and $N_{f,b2z}=153/19=8.053$ in this SU(3) theory, the
maximal value of $\Delta_f$ in the interval $I_{IRZ}$ is 
\beq
(\Delta_f)_{\rm max} = \frac{321}{38} = 8.447 \quad {\rm for} \ {\rm SU}(3), \ 
N_f \in I_{IRZ} \ .  
\label{Delta_max_su3_irz}
\eeq
Therefore, these ratios suggest that the small-$\Delta_f$ expansion may be
reasonably reliable in most of this interval, $I_{IRZ}$ and the associated
non-Abelian Coulomb phase.


\subsection{Calculation in the LNN Limit and Comparison with Conventional 
Calculation}
\label{betaprime_lnn_section}

For theories having gauge the group $G={\rm SU}(N_c)$ with $N_f$ fermions in
the fundamental representation of this group, i.e., $R=F$, it is of interest to
consider the limit
\beqs
& & N_c \to \infty \ , \quad N_f \to \infty \cr\cr
& & {\rm with} \ r \equiv \frac{N_f}{N_c} \ {\rm fixed \ and \ finite}  \cr\cr
& & {\rm and} \ \ \xi(\mu) \equiv \alpha(\mu) N_c \ {\rm is \ a \ 
finite \ function \ of} \ \mu \ . 
\cr\cr
& &
\label{lnn}
\eeqs
We will use the symbol $\lim_{LNN}$ for this limit, where ``LNN'' stands
for ``large $N_c$ and $N_f$'' (with the constraints in Eq. (\ref{lnn})
imposed).  In this LNN ('t Hooft-Veneziano) limit we define the quantities 
\beq
x = \lim_{LNN} \frac{g^2 N_c}{16\pi^2} = \frac{\xi}{4\pi} \ , 
\label{xlnn}
\eeq
\beq
r_{b1z} = \lim_{LNN} \frac{N_{f,b1z}}{N_c} \ , 
\label{rb1zdef}
\eeq
and
\beq
r_{b2z} = \lim_{LNN} \frac{N_{f,b2z}}{N_c} \ , 
\label{rb2zdef}
\eeq
with values 
\beq
r_{b1z} = \frac{11}{2} =5.5
\label{rb1z}
\eeq
and
\beq
r_{b2z} = \frac{34}{13}=2.615  \ . 
\label{rb2z}
\eeq
(to the indicated floating-point accuracy).  
With $I_{IRZ}$ being $N_{f,b2z} < N_f < N_{f,b1z}$, the corresponding 
interval in the ratio $r$ is 
\beq
I_{IRZ,r}: \quad \frac{34}{13} < r < \frac{11}{2}, \ i.e., 
\ 2.615 < r < 5.5 
\label{intervalr}
\eeq
We define the scaled scheme-independent expansion parameter for the LNN limit
\beq
\Delta_r \equiv \frac{\Delta_f}{N_c} = r_{b1z}-r = \frac{11}{2}-r \ . 
\label{deltar}
\eeq
and 
\beq
r_c = \lim_{LNN} \frac{N_{f,cr}}{N_c} \ . 
\label{rc}
\eeq

After these preliminaries, we now proceed to calculate the scheme-independent
expansion of $\beta'_{IR}$ in this LNN limit (\ref{lnn}), and
compare with the conventional calculation of this quantity.  The beta function
that is finite in this LNN limit is
\beq
\beta_\xi = \frac{d\xi}{dt} = \lim_{LNN} \beta N_c \ , 
\label{betaxi}
\eeq
where $\xi = \lim_{LNN} \alpha N_c$ was defined in Eq. (\ref{lnn}). This has
the series expansion
\beq
\beta_\xi \equiv \frac{d\xi}{dt}
= -8\pi x \sum_{\ell=1}^\infty \hat b_\ell x^\ell \ , 
\label{betaxiseries}
\eeq
where 
\beq
\hat b_\ell = \lim_{LNN} \frac{b_\ell}{N_c^\ell} \ . 
\label{bellhat}
\eeq
The $\hat b_\ell$ are listed for reference in Appendix \ref{bellhatappendix}. 

Since the derivative $d\beta_\xi/d\xi$ satisfies the relation 
\beq
\frac{d \beta_\xi}{d\xi} = \frac{d\beta}{d\alpha} \equiv \beta' \ , 
\label{dbetarelation}
\eeq
it follows that $\beta'$ is finite in the
LNN limit (\ref{lnn}).  In terms of the variable $x$ 
defined in Eq. (\ref{xlnn}), we have
\beq
\beta' = -2\sum_{\ell=1}^\infty (\ell+1) \hat b_\ell \, x^\ell \ . 
\label{betaprime_lnn}
\eeq

Because $\beta'_{IR}$ is scheme-invariant and is finite in the LNN limit, one
is motivated to calculate the LNN limit of the scheme-independent expansion
(\ref{betaprime_delta}).  For this purpose, in addition to the rescaled
quantities $\Delta_r$ defined in Eq. (\ref{deltar}), we define the rescaled
coefficient
\beq
\hat d_n = N_c^n \, d_n \ , 
\label{dnhat}
\eeq
which is finite in the LNN limit.  Then each term 
\beq
\lim_{LNN} d_n \Delta_f^n= (N_c^n d_n)\Big ( \frac{\Delta_f}{N_c} \Big )^n = 
\hat d_n \Delta_r^n 
\label{finiteproduct}
\eeq
is finite in this limit. 
Thus, writing $\lim_{LNN} \beta'_{IR}$ as $\beta'_{IR,LNN}$, we have 
\beqs
& & \beta'_{IR,LNN} = \sum_{n=1}^\infty d_n \Delta_f^n = 
\sum_{n=1}^\infty \hat d_n \Delta_r^n \ . \cr\cr
& &  
\label{betaprime_ir_lnn}
\eeqs
We denote the value of $\beta'_{IR,LNN}$ obtained from 
this series calculated to order $O(\Delta_f^p)$ as 
$\beta'_{IR,LNN,\Delta_f^p}$. 

From Eqs. (\ref{d1})-(\ref{d4}), we find that the approach to the LNN limits
for $\hat d_n$ involves correction terms that vanish like $1/N_c^2$.  This is
the same property that was found in \cite{bc,lnn} and, in the same way, it
means that the approach to the LNN limit for finite $N_c$ and $N_f$ with fixed
$r=N_f/N_c$ is rather rapid, as discussed in \cite{lnn}.  We calculate 
$\hat d_1=0$ and
\beqs
\hat d_2 & = \frac{2^4}{3^2 \cdot 5^2} \cr\cr
         & = & 0.0711111 \ , 
\label{d2hat_lnn}
\eeqs
\beqs
\hat d_3 & = & \frac{416}{3^3 \cdot 5^4} \cr\cr
& = & 2.465185 \times 10^{-2} \ , 
\label{d3hat_lnn}
\eeqs
and
\beqs
\hat d_4 &=& \frac{5868512}{3^5 \cdot 5^{10}}
-\frac{5632}{3^4 \cdot 5^6} \, \zeta_3 \cr\cr
& = & -(2.876137 \times 10^{-3}) \ . 
\label{d4hat_lnn}
\eeqs
Thus, numerically
\beqs
\beta'_{IR,LNN} &=& \Delta_r^2\Big [0.07111 + 
(2.4652 \times 10^{-2})\Delta_r \cr\cr
&-& (2.8761 \times 10^{-3})\Delta_r^2 \Big ] \ .
\label{betaprime_numerical}
\eeqs

We may again calculate ratios of successive magnitudes of these coefficients to
get a rough estimate of the range over which the small-$\Delta_r$ expansion is
reliable in this LNN limit.  We find
\beq
\frac{\hat d_2}{\hat d_3} = 2.885
\label{d2hat_over_d3hat}
\eeq
and
\beq
\frac{\hat d_3}{|\hat d_4|} = 8.571
\label{d3hat_over_d4hat}
\eeq
For $r \in I_{IRZ,r}$, the maximal value of $\Delta_r$ is
\beq
(\Delta_r)_{\rm max} = \frac{75}{26} = 2.885 \quad {\rm for} \ r \in I_{IRZ,r}
\ .
\label{Deltar_max_irz}
\eeq
Therefore, these LNN ratios suggest, in agreement with our analysis for SU(3)
and $R=F$, that the small-$\Delta_r$ expansion may be reasonably reliable over
much of the interval $I_{IRZ,r}$.

It is useful to compare these scheme-independent calculations of
$\beta'_{IR,LNN}$ with the results of conventional $n$-loop
calculations, denoted $\beta'_{IR,n\ell,LNN}$. 
These derivatives are computed
from the $n$-loop truncation of the series in Eq. (\ref{betaprime_lnn}).  As a
special case of our remark below Eq.  (\ref{betaprime_series}), we note that in
calculating the $n$-loop truncation of the series (\ref{betaprime_lnn}) at the
IR zero of the beta function, for $n \ge 3$, one uses the property that
\beq
\sum_{\ell=1}^n \hat b_\ell \, x_{IR,n\ell}^{\ell-1}=0 \ , 
\label{xir_nloop_condition}
\eeq
to eliminate the highest-loop term $\hat b_n x_{IR}^{n-1}$, expressing it as 
$\hat b_n x_{IR}^{n-1}=-\sum_{\ell=1}^{n-2} \hat 
b_\ell \, x_{IR,n\ell}^{\ell-1}$. 
The two-loop result for $x_{ir}$ is  
\beq
x_{IR,2\ell}=\frac{11-2r}{13r-34} \quad {\rm for} \ r \in I_{IRZ,r} \ .
\label{xir_2loop}
\eeq
The resultant two-loop for $\beta'_{IR}$ is 
\beq
\beta'_{IR,2\ell} = \frac{2(11-2r)^2}{3(13r-34)} \ . 
\label{betaprime_ir_2loop}
\eeq
Both $x_{IR,2\ell}$ and $\beta'_{IR,2\ell}$ 
are scheme-independent.  However, the higher-loop expressions for 
these quantities at loop level $n \ge 3$ do not preserve the
scheme-independence of the exact $\beta'_{IR}$.  
Let us define the polynomials (see Eqs. (3.9) and (2.26) in \cite{lnn})
\beq
C_{3\ell} = -52450+41070r-7779r^2+448r^3
\label{c3ell}
\eeq
and
\beq
D_{3\ell} = -2857+1709r-112r^2 \ , 
\label{d3ell}
\eeq
both of which are positive for $r \in I_{IRZ,r}$.  The three-loop value of the
IR zero of the beta function in the LNN limit, computed in the 
$\overline{\rm MS}$ scheme, is \cite{lnn}
\beq
x_{IR,3\ell}=\frac{3[-3(13r-34) + \sqrt{C_{3\ell}} \ ]}{D_{3\ell}} \ . 
\label{xiir_3loop_explicit}
\label{xir_3loop}
\eeq
We calculate the three-loop result for $\beta'_{IR}$, or equivalently 
the anomalous dimension of ${\rm Tr}(F_{\mu\nu}F^{\mu\nu})$, 
in the LNN limit, again in the $\overline{\rm MS}$ scheme, to be
\begin{widetext}
\beqs
\beta'_{IR,3\ell} & = & \frac{2\Big [ -3(13r-34)+\sqrt{C_{3\ell}} \ \Big ]}
{D_{3\ell}^2} \, \Big [ -52450+41070r-7779r^2+448r^3 
-3(13r-34)\sqrt{C_{3\ell}} \ \Big ] \ . 
\label{betaprime_3loop_ir_lnn}
\eeqs
\end{widetext}
We compute the four-loop result $\beta'_{IR,4\ell}$ in
this scheme in a similar manner.  In Table \ref{betaprime_values_lnn} we list
the numerical values of these conventional $n$-loop calculations in comparison
with our scheme-independent results calculated to $O(\Delta_f^p)$ for 
$2 \le n \le 4$ and $1 \le p \le 3$. We see that, especially for $r$ values 
in the upper part of the interval $I_{IRZ,r}$, the results are rather close,
and, furthermore, that, as expected, for a given $r$, the higher the loop level
$n$ and the truncation order $p$ in the respective calculations of
$\beta'_{IR,n\ell}$ in the $\overline{\rm MS}$ scheme and the
scheme-independent $\beta'_{IR,\Delta_f^p}$, the better the agreement between
these two results.  All of the entries
shown in Table \ref{betaprime_values_lnn} have $\beta'_{IR} < 3$ 
except for the two-loop values $\beta'_{IR,2\ell}$ for $r=3.0$
and $r=2.8$ which are 3.333 and 8.100, respectively.


\subsection{Calculation of the $d_n$ to $O(\Delta_f^4)$ for 
$G={\rm SU}(N_c)$ and $R=adj$}
\label{betaprime_adj_section}

It is worthwhile to compare our results obtained for $G={\rm SU}(N_c)$ with
$N_f$ fermions in the fundamental representation to the case in which the
fermions
are in the adjoint representation, denoted as $adj$ for short. In this case,
the general expressions for $N_{f,b1z}$ and $N_{f,b2z}$ are 
\beq
N_{f,b1z} = \frac{11}{4} = 2.75 \quad {\rm for} \ R = adj 
\label{nfb1z_adj}
\eeq
and
\beq
N_{f,b2z} = \frac{17}{16} = 1.0625 \quad {\rm for} \ R = adj \ , 
\label{nfb2z_adj}
\eeq
so the interval $I_{IRZ}$ only contains the single integer value
$N_f=2$. 

For this theory, our general expressions (\ref{d2}) and (\ref{d3}) reduce to 
pure numbers, independent of $N_c$:
\beq
d_{2,{\rm SU}(N_c),adj} = \frac{2^4}{3^4} = 0.19753 \ ,  
\label{d2adj}
\eeq
\beq
d_{3,{\rm SU}(N_c),adj} = \frac{2^8}{3^7} = 0.11706 \ ,  
\label{d3adj}
\eeq
For $d_4$ we calculate 
\beq
d_{4,{\rm SU}(N_c),adj} = \frac{46871N_c^2+ 85248}{2^2 \cdot 3^{12}N_c^2} \ .
\label{d4adj}
\eeq
This coefficient $d_{4,{\rm SU}(N_c),adj}$ is manifestly positive
and has the large-$N_c$ limit
\beqs
& & \lim_{N_c \to \infty} d_{4,{\rm SU}(N_c),adj} = \frac{46871}
{2^2 \cdot 3^{12}} = 0.022049   \cr\cr
& & 
\label{d4adj_ncinfinity}
\eeqs
In contrast to our results for the $d_{n,{\rm SU}(N_c),F}$, here all of the 
coefficients $d_{n,{\rm SU}(N_c),adj}$ that we have calculated, for 
$1 \le n \le 4$, are positive.  
These signs are recorded in Table \ref{signs}.  

With these coefficients, one can again compute ratios to obtain a crude idea of
the region over which the small-$\Delta_f$ series expansion is reliable. We
have 
\beq
\frac{d_{2,{\rm SU}(N_c),adj}}{d_{3,{\rm SU}(N_c),adj}} = 1.687
\label{d2_over_d3_adj}
\eeq
and, taking the large-$N_c$ limit for simplicity, 
\beq
\lim_{N_c \to \infty} 
\frac{d_{3,{\rm SU}(N_c),adj}}{d_{4,{\rm SU}(N_c),adj}} = 5.309
\label{d3_over_d4_adj}
\eeq
These ratios are consistent with the inference that the small-$\Delta_f$
expansion may again be reasonably accurate in the interval $I_{IRZ}$ and for
the corresponding value $N_f=2$ in this theory.


\section{Analysis of Scheme-Independent Expansion Coefficients for 
$\gamma_{\bar\psi\psi,IR}$}
\label{kappa_section}


\subsection{Review of Calculation to $O(\Delta_f^3)$ for General $G$ and $R$}
\label{kappa_review_section}

We consider the (gauge-invariant) flavor-nonsinglet ($fns$) and 
flavor-singlet ($fs$) bilinear fermion operators 
\beq
J_{0,fns} = \sum_{j,k=1}^{N_f} \bar\psi_j (T_b)_{jk} \psi_k  \ , 
\label{j0fns}
\eeq
where here $T_b$ with $b=1,...,N_f^2-1$ is an generator of the global flavor 
group SU($N_f$), and 
\beq
J_{0,fs} = \sum_{j=1}^{N_f} \bar\psi_j \psi_j \ .
\label{j0fs}
\eeq
We will often suppress the flavor indices and write these simply as 
$\bar\psi T_b \psi$ and $\bar\psi \psi$.  These have the same anomalous 
dimension (e.g., \cite{gracey_gammatensor}), which we denote as 
$\gamma_{\bar\psi\psi}$.  (Thus, one may simply consider the operator 
$\bar\psi_j \psi_j$ with no sum on $j$, but here we shall refer to 
$J_{0,fns}$ and $J_{0,fs}$.) The operator $J_{0,fns}$ has the chiral 
decomposition 
$\bar\psi T_b \psi = \bar\psi_L T_b \psi_R + \bar\psi_R T_b \psi_L$. 
Hence, in the non-Abelian Coulomb phase where the flavor symmetry is
(\ref{gfl}), one may regard the $T_b$ in the term $\bar\psi_L T_b \psi_R$ 
acting to the right as an element of ${\rm SU}(N_f)_R$ and acting to the 
left as an element of ${\rm SU}(N_f)_L$.

The usual series expansion of $\gamma_{\bar\psi\psi}$ in powers of $\alpha$, or
equivalently, $a$, is 
\beq
\gamma_{\bar\psi\psi} = \sum_{\ell=1}^\infty c_\ell \, a^\ell \ , 
\label{gamma_aseries}
\eeq
where $c_\ell$ is the $\ell$-loop coefficient.  For general $G$ and $R$ the
coefficients $c_\ell$ have been calculated up to $\ell=4$ loop level \cite{c4}
(earlier work includes \cite{jmearlier}) and for the special case 
$G={\rm SU}(3)$ and $R=F$, $c_5$ has been calculated \cite{c5su3}.  The
scheme-independent expansion of $\gamma_{\bar\psi\psi}$ can be written as
\beq
\gamma_{\bar\psi\psi,IR} = \sum_{n=1}^\infty \kappa_n \, \Delta_f^n \ . 
\label{gamma_psibarpsi_series}
\eeq
We denote the truncation of this sum to maximal power $n=p$ as 
$\gamma_{\bar\psi\psi,IR,\Delta_f^p}$. 
For a general asymptotically free vectorial gauge theory with gauge group $G$
and $N_f$ fermions in an arbitrary representation $R$, the coefficients
$\kappa_n$ were given in \cite{gtr} up to order $n=3$, yielding the 
expansion of $\gamma_{\bar\psi\psi,IR}$ to order $\Delta_f^3$. 
For reference, we display the $\kappa_n$ coefficients from Ref. \cite{gtr} 
(with the denominator factor $D$ given in Eq. (\ref{d})):
\beq
\kappa_1 = \frac{8T_fC_f}{C_AD} \ , 
\label{kappa1}
\eeq
\beq
\kappa_2 = \frac{4T_f^2C_f(5C_A+88C_f)(7C_A+4C_f)}{3C_A^2 D^3} \ , 
\label{kappa2}
\eeq
and 
\begin{widetext}
\beqs
\kappa_3 &=& \frac{4T_fC_f}{3^4 C_A^4 D^5} \bigg [ 
-55419T_f^2C_A^5 + 432012T_f^2C_A^4C_f 
+ 5632T_f^2 C_f \, \frac{d_A^{abcd}d_A^{abcd}}{d_A} \, (-5+132\zeta_3) \cr\cr
&+& 
16C_A^3 \bigg ( 122043T_f^2 C_f^2 + 6776 \, \frac{d_R^{abcd}d_R^{abcd}}{d_A} 
\, (-11+24\zeta_3) \bigg ) \cr\cr
&+& 704C_A^2 \bigg ( 1521 T_f^2 C_f^3 + 112 T_f \, 
\frac{d_R^{abcd}d_A^{abcd}}{d_A} \, (4-39\zeta_3) 
+ 242C_f \, \frac{d_R^{abcd}d_R^{abcd}}{d_A} \, (-11+24\zeta_3) \bigg ) 
\cr\cr
&+& 32T_fC_A \bigg ( 53361T_fC_f^4 - 3872 C_f 
\, \frac{d_R^{abcd}d_A^{abcd}}{d_A} \, (-4+39\zeta_3) 
+ 112T_f \, \frac{d_A^{abcd}d_A^{abcd}}{d_A} \, (-5+132\zeta_3) \bigg ) 
\bigg ] \ .  
\label{kappa3}
\eeqs
\end{widetext} 
%


\subsection{Evaluation of $\kappa_n$ for $G={\rm SU}(N_c)$ and $R=F$}
\label{kappa_sun_section}

For the case where the $N_f$ fermions are in the representation 
$R=F$, these results (\ref{kappa1})-(\ref{kappa3}) from \cite{gtr} 
take the following forms:
\beq
\kappa_{1,{\rm SU}(N_c),F} = \frac{4(N_c^2-1)}{N_c(25N_c^2-11)} \ , 
\label{kappa1f}
\eeq
\beq
\kappa_{2,{\rm SU}(N_c),F} = \frac{4(N_c^2-1)(9N_c^2-2)(49N_c^2-44)}
{3N_c^2(25N_c^2-11)^3} \ , 
\label{kappa2f}
\eeq
and
\begin{widetext}
\beqs
\kappa_{3,{\rm SU}(N_c),F} &=& 
\frac{8(N_c^2-1)}{3^3N_c^3(25N_c^2-11)^5} \, 
\Big [ 274243N_c^8-455426N_c^6-114080N_c^4+47344N_c^2+35574 \cr\cr
&-& 4224N_c^2(4N_c^2-11)(25N_c^2-11)\zeta_3 \Big ] \ . 
\label{kappa3f}
\eeqs
\end{widetext}
We find that these coefficients $\kappa_{n,{\rm SU}(N_c),F}$ with 
$1 \le n \le 3$ are positive-definite for all physical $N_c \ge 2$. This is
obvious for $n=1,2$, and an examination of the polynomial in square brackets in
Eq. (\ref{kappa3f}), of degree 8 in $N_f$, proves the result for $n=3$. 


\subsection{Calculation of $\kappa_n$ Coefficients to $O(\Delta_f^4)$ for 
$G={\rm SU}(3)$ and $R=F$}
\label{kappa_su3_section}

For comparison with the $\kappa_n$ with other values of $N_c$, we 
recall our calculation of the $\kappa_n$ to order $n=4$, i.e., to order 
$O(\Delta_f^4)$ in \cite{gsi}.  We found 
\beq
\kappa_{{\rm SU}(3),F,1} = \frac{16}{3 \cdot 107} = 4.9844 \times 10^{-2} \ , 
\label{kappa1j0su3f}
\eeq
\beq
\kappa_{{\rm SU}(3),F,2} = 
\frac{125452}{(3 \cdot 107)^3} = 3.7928 \times 10^{-3} \ , 
\label{kappa2j0su3f}
\eeq
\beq
\kappa_{{\rm SU}(3),F,3} = \frac{972349306}{(3 \cdot 107)^5} -
\frac{140800}{3^3 \cdot (107)^4} \, \zeta_3 = 2.3747 \times 10^{-4} \ , 
\label{kappa3j0su3f}
\eeq
\beqs
\kappa_{{\rm SU}(3),F,4} & = & \frac{33906710751871}{2^2 (3 \cdot 107)^7}
-\frac{1684980608}{3^5 \cdot (107)^6} \, \zeta_3 \cr\cr
& + & \frac{59840000}{(3 \cdot 107)^5} \, \zeta_5 \cr\cr
& = & 3.6789 \times 10^{-5} \ .
\label{kappa4j0su3f}
\eeqs

In Ref. \cite{gtr} the ratio test was applied to the first three coefficients,
$\kappa_{{\rm SU}(3),F,n}$, $n=1,2,3$ and the excellent convergence was noted.
Here, using our calculation of $\kappa_{{\rm  SU}(3),F,4}$ in \cite{gsi}, we
calculate the next ratio,  
$\kappa_{{\rm  SU}(3),F,3}/\kappa_{{\rm SU}(3),F,4}$. We have 
\beq
\frac{\kappa_{{\rm SU}(3),F,1}}{\kappa_{{\rm SU}(3),F,2}} = 13.142
\label{kappa1_over_kappa2_su3f}
\eeq
\beq
\frac{\kappa_{{\rm SU}(3),F,2}}{\kappa_{{\rm SU}(3),F,3}} = 15.972
\label{kappa2_over_kappa3_su3f}
\eeq
and
\beq
\frac{\kappa_{{\rm SU}(3),F,3}}{\kappa_{{\rm SU}(3),F,4}} = 6.455
\label{kappa2_over_kappa4_su3f}
\eeq
Since the maximal value of $\Delta_f$ in the interval $I_{IRZ}$ is 8.447 (see
Eq. (\ref{Delta_max_su3_irz})), these ratios suggest, as noted in \cite{gtr}
and in agreement with our earlier calculation of coefficient ratios for
$\beta'_{IR}$, that the small-$\Delta_f$ expansion may be reasonably reliable
over much of the interval $I_{IRZ}$.

The positivity of the $\kappa_{{\rm SU}(3),F,n}$ for $1 \le n \le 3$ is in
agreement with our more general positivity results given above, and, as we
noted in \cite{gsi}, we also found that $\kappa_{{\rm SU}(3),F,4}$ is 
positive.  These signs are recorded in Table \ref{signs}.  The positivity
of all of these coefficients played an important role in our analysis in
\cite{gsi} because it meant that for a given value of $N_f$, or equivalently,
$\Delta_f$, the value of $\gamma_{\bar\psi\psi}$ calculated to $O(\Delta_f^n)$,
denoted $\gamma_{\bar\psi\psi,\Delta_f^n}$, is a monotonically increasing
function of $n$ over the full range $1 \le n \le 4$ that we calculated. We then
conjectured that this positivity would be true for all $n$, i.e., we
conjectured that $\kappa_n > 0$ for all $n \ge 1$.  Assuming the
validity of this conjecture, we then computed the extrapolation to 
$n \to \infty$ for an exact $\gamma_{\bar\psi\psi,IR}$ in the SU(3) theory 
with $R=F$.  A
generalization of our conjecture in \cite{gsi} that is motivated by our present
results is that, in the notation of 
Eqs. (\ref{kappa1j0su3f})-(\ref{kappa4j0su3f}), 
$\kappa_{n,{\rm SU}(N_c),F} > 0$ for all $n \ge 1$ and all 
$N_c \ge 2$.  Importantly, in \cite{gsi} we noted that, if this monotonicity
property holds, then, combining it with the upper bound 
$\gamma_{\bar\psi\psi,IR} < 2$, one would infer that if $\gamma_{IR}$ 
saturates its upper bound (\ref{gamma_upperbound}) as $N_f$ decreases and 
passes through the value $N_{f,cr}$ at the lower end of the non-Abelian 
Coulomb phase, it would follow from our extrapolated values of 
$\gamma_{\bar\psi\psi,IR}$ that $8 < N_{f,cr} < 9$.   Here one must mention 
the caveat that it is not known if, in fact, $\gamma_{IR}$ saturates its 
upper bound in this way as $N_f \searrow N_{f,cr}$. Indeed, the nature of the
transition as $N_f$ decreases through $N_{f,cr}$ has not been definitely
established. Analyses via Schwinger-Dyson equations suggested that, as 
$N_f \nearrow N_{f,cr}$ from within the phase with confinement and chiral 
symmetry breaking, the fermion condensate $\langle \bar\psi\psi \rangle$ 
could vanish with an essential zero \cite{miranskyscaling}. 
Some insight into this may be derived from the known results in SQCD. 
In SQCD, as noted above, the upper bound is $\gamma_{\bar\psi\psi,IR} < 1$ and
is saturated at the lower end of the non-Abelian Coulomb phase
\cite{nsvz,seiberg} . 

In the case $G={\rm SU}(3)$ and $R=F$, one of the major values of the five-loop
calculation of $\gamma_{\bar\psi\psi,IR}$ in \cite{flir} and the
scheme-independent calculations of $\gamma_{\bar\psi\psi,IR}$ to order
$\Delta_f^3$ in \cite{gtr} and to order $\Delta_f^4$ in \cite{gsi}, with the
additional analysis here, is the comparison of these results with fully
nonperturbative lattice measurements of this anomalous dimension
\cite{lgtreviews}.  (Since our discussion here is on the operator
$\bar\psi\psi$ and the gauge group SU(3), when there is no
danger of confusion, we omit these subscripts in the $\bar\psi\psi$
anomalous dimension.)  A number of lattice groups have obtained data and
carried out analyses of these data for the SU(3) theory with $N_f=12$ fermions
with $R=F$. These groups have reported the following values: $\gamma_{IR} =
0.414 \pm 0.016$ \cite{lsd}, $\gamma_{IR} \sim 0.35$ \cite{degrand},
$\gamma_{IR} \simeq 0.4$ \cite{latkmi}, $\gamma_{IR} = 0.27(3)$ \cite{ah1},
$\gamma_{IR} \simeq 0.25$ \cite{ah2}, $\gamma_{IR} = 0.235(46)$ \cite{lmnp},
and $0.2 \lsim \gamma \lsim 0.4$ \cite{kuti}. (For comparative discussions of
these different results and estimates of overall uncertainties, the reader is
advised to consult the reviews in \cite{lgtreviews} and the original papers
\cite{lsd}-\cite{kuti}.)  As we noted in \cite{gsi}, our value
$\gamma_{IR,\Delta^4}=0.338$ and our extrapolated $\gamma_{IR} = 0.40$ are
consistent with this range of lattice measurements, taking into account the
different methods of lattice data analysis used and are somewhat higher than
the five-loop value $\gamma_{IR,5\ell}=0.255$ from the conventional $\alpha$
series that we obtained in \cite{flir}. The $\gamma_{IR,5\ell}=0.255$ value in
\cite{flir} is in very good agreement with the measured values of $\gamma_{IR}$
reported in \cite{ah1}-\cite{lmnp}.

There have also been lattice studies of the SU(3) theory with $N_f=10$
\cite{lsdnf10} and $N_f=8$ \cite{latkminf8,lsdnf8,lgtreviews}.  For the SU(3)
theory with $N_f=10$ fermions, our scheme-independent calculation presented in
\cite{gsi} and discussed further here gives
$\gamma_{\bar\psi\psi,IR,\Delta_f^4} = 0.615$ and our extrapolation to infinite
order in the $\Delta_f$ expansion yields $\gamma_{\bar\psi\psi,IR} = 0.95(6)$,
consistent with estimates that $\gamma_{\bar\psi\psi,IR} \sim 1$ from lattice
studies \cite{lsdnf10,lgtreviews}. In the SU(3) theory (with $N_f$ fermions in
the representation $R=F$), the lower end of $I_{IRZ}$ occurs at
$N_{f,b2z}=8.047$, but one may still formally consider the results of the
small-$\Delta_f$ expansion evaluated at $N_f=8$.  In this case we obtain
$\gamma_{IR,\Delta_f^p}=0.424, \ 0.698, \ 0.844, \ 1.04$ for $1 \le p \le 4$.
These are again consistent with the rough estimates $\gamma_{\bar\psi\psi,IR}
\sim 1$ from lattice studies \cite{latkminf8,lsdnf8,lgtreviews}.  There is not
yet a consensus on the value of $N_{f,cr}$ from lattice studies
\cite{lgtreviews}.  In this context, one should keep in mind that for 
$N_f < N_{f,cr}$, there is spontaneous chiral symmetry breaking, so the IR 
zero of the beta function is only approximate, since the theory flows away from
this value as the fermions gain dynamical mass and are integrated out, leaving
a pure gluonic low-energy effective field theory. For such a theory, the
quantity extracted from either continuum or lattice analyses as
$\gamma_{\bar\psi\psi,IR}$ is only an effective anomalous dimension that
describes the renormalization-group behavior as the theory is flowing near to
the approximate zero of the beta function.


\subsection{Evaluation of $\kappa_{n,{\rm SU}(N_c),R}$ to $O(\Delta_f^3)$ for 
$R=adj$}
\label{kappa_adj_section}

In the case $R=adj$, the general results in \cite{gtr} reduce as follows:
\beq
\kappa_{1,{\rm SU}(N_c),adj}= \frac{4}{3^2} = 0.4444 \ , 
\label{kappa1_adj}
\eeq
\beq
\kappa_{2,{\rm SU}(N_c),adj}= \frac{341}{2 \cdot 3^6} = 0.23388 \ ,  
\label{kappa2_adj}
\eeq
\beq
\kappa_{3,{\rm SU}(N_c),adj} = \frac{61873N_c^2-42624}
{2^3 \cdot 3^{10}N_c^2}\ . 
\label{kappa3_adj}
\eeq
This is positive for all physical $N_c$ and has the large-$N_c$ limit
\beq
\lim_{N_c \to \infty} \kappa_{3,{\rm SU}(N_c),adj} = \frac{61873}
{2^3 \cdot 3^{10}} = 0.130978
\label{kappa3_adj_largeNc}
\eeq
The positive signs of these $\kappa_{n,{\rm SU}(N_c),adj}$ coefficients are
recorded in Table \ref{signs}.


\subsection{Comparison of Scheme-Independent Calculation of 
$\gamma_{\bar\psi\psi,IR}$ with Conventional Calculations}
\label{kappa_comparison_section}

It is of considerable interest to compare the results obtained in \cite{gtr}
for the scheme-independent expansion of $\gamma_{\bar\psi\psi,IR}$ to order 
$O(\Delta_f^3)$ (using calculations of the $b_n$ to $n=4$ loop order and 
$c_n$ to $n=3$ loop order) with results obtained previously
with the conventional calculation of the $n$-loop
$\gamma_{\bar\psi\psi,IR,n\ell}$ in powers of the $n$-loop $\alpha_{IR,n\ell}$
in \cite{bvh} (using calculations of the $b_n$ and $c_n$ up to $n=4$ loop 
order). Here and below, for specific calculations we take the gauge group to be
SU($N_c$) with various values of $N_c$.  
For notational brevity, in this section we will often leave the subscript
$\bar\psi\psi$ implicit on these and other quantities and thus write 
$\gamma_{IR} \equiv \gamma_{\bar\psi\psi,IR}$, 
$\gamma_{IR,n\ell} \equiv \gamma_{\bar\psi\psi,IR,n\ell}$, 
$\kappa_n \equiv \kappa_{\bar\psi\psi,n}$, etc. in this and the next section. 
Since $\gamma_{IR,n\ell}$ is scheme-dependent beyond the lowest order, one
must choose a scheme for this comparison.  Here we choose the widely used
$\overline{\rm MS}$ scheme, for which $b_3$ and $b_4$ and $c_n$ for
$2 \le n\le 4$ were calculated for a general gauge group $G$ and fermion
representation $R$ \cite{b3,b4,b4p} \cite{c4}.  
In the special case of $G={\rm SU}(3)$ and $R=F$, using the recent 
calculations of the five-loop coefficients $b_5$ and $c_5$ in the 
$\overline{\rm MS}$ scheme, we computed $\gamma_{IR,n\ell}$ up to
$n=5$ loop level \cite{flir} in this $\overline{\rm MS}$ scheme and performed a
scheme-independent calculation up to order $\Delta_f^4$ \cite{gsi}.  For this
special case we compared the results obtained via these two different
approaches.  Here we carry out a similar comparison for other SU($N_c$) 
theories. The scheme-independent expansion of $\gamma_{IR}$ 
has the form (\ref{gamma_psibarpsi_series}). 
We denote the value of $\gamma_{IR}$ obtained from this series calculated to 
order $O(\Delta_f^p)$ as $\gamma_{IR,\Delta_f^p}$ 

As discussed above, our discussion is restricted to the interval $I_{IRZ}$ of
values of $N_f$, given in Eq. (\ref{nfinterval}), for which the
(scheme-independent) two-loop beta function has an IR zero.  Using the results
for the lower and upper ends of this interval, $N_{f,b2z}$ and $N_{f,b1z}$ from
Eqs. (\ref{nfb1z}) and (\ref{nfb2z}), one has, for $(N_{f,b1z},N_{f,b2z})$, the
respective values $(5.55,11)$, $(8.05,16.5)$, and $(10.61,22)$ for $N_c=2, \ 3,
\ 4$ \cite{nfintegral}, and hence the physical intervals $I_{IRZ}$ with
integral $N_f$: $6 \le N_f \le 10$ for SU(2), $9 \le N_f \le 16$ for SU(3), and
$11 \le N_f \le 21$ for SU(4). Our results for these three
illustrative values of $N_c$ are listed in Table \ref{gamma_values}. For 
the special case $N_c=3$, we have carried these calculations one order higher,
namely to five-loop level and to order $\Delta_f^4$ in \cite{flir,gsi}.

Since the calculation of $\kappa_n$ and the resultant $\gamma_{IR,\Delta_f^n}$
uses information from the $(n+1)$-loop beta function from (\ref{beta}) and the
$n$-loop expansion of $\gamma$ in (\ref{gammaseries}), it is natural to compare
the (SI) $\gamma_{IR,\Delta_f^n}$ with the (SD) $\gamma_{IR,n'\ell}$ for $n'=n$
and $n'=n+1$.  Since
$\gamma_{IR,\Delta_f^n}$ includes $n$-loop information about
$\gamma_{IR,n\ell}$, one would expect the closest agreement between
$\gamma_{IR,\Delta_f^n}$ and $\gamma_{IR,n\ell}$, and our results confirm this
expectation.  In the upper and middle part of the interval $I_{IRZ}$ for a
given $N_c$, we find that $\gamma_{IR,\Delta_f^n}$ is slightly larger than
$\gamma_{IR,3\ell}$, with the difference increasing as $N_f$ decreases below
$N_{f,b1z}$, i.e., as $\Delta_f$ increases.

We recall
the upper bound (\ref{gamma_upperbound}) that applies at an IRFP in the 
non-Abelian Coulomb phase, based on the scale invariance and inferred conformal
invariance in this phase. The bound (\ref{gamma_upperbound}) also applies, for
a different reason, in the phase with confinement and spontaneous chiral
symmetry breaking; in that phase it is a consequence of the physical
requirement that the momentum-dependent dynamically generated effective fermion
mass
\beq
m(k) \sim \Lambda \Big ( \frac{\Lambda}{k} \Big )^{2-\gamma_{IR}}
\label{mk}
\eeq
must satisfy the constraint $\lim_{k \to \infty} m(k)=0$, where $k$ is the
Euclidean momentum.  In the upper and middle parts of the interval $I_{IRZ}$ in
the NACP, the values of $\gamma_{IR,n\ell}$ calculated in the conventional
series expansion in powers of $\alpha_{IR,n\ell}$ obey this upper bound.
However, for a given $N_c$, toward the lower end of the respective intervals
$I_{IRZ}$, the IR coupling $\alpha_{IR,n\ell}$ become too large for the
perturbative calculations to be applicable, and some resultant values of the
anomalous dimensions exceed the bound (\ref{gamma_upperbound}). This occurs for
the scheme-independent two-loop values $\gamma{IR,2\ell}$ for $N_f=6, \ 7$ if
$N_c=2$; for $N_f=9, \ 10$ if $N_c=3$, and for $11 \le N_f \le 14$ if
$N_f=4$. In these cases, since it is not clear that the higher-order values
$\gamma_{IR,n\ell}$ are reliable, we leave them unlisted (u), as we did in
\cite{bvh}.

From these calculations and the entries in Table \ref{gamma_values}, one of the
important advances achieved by the scheme-independent $\Delta_f$ expansion is
evident, namely that the values of $\gamma_{IR,\Delta_f^p}$ with $1 \le p \le
3$ (and, for SU(3) also $p=4$ in \cite{gsi}) that we calculate via this method
obey the upper bound (\ref{gamma_upperbound}) throughout all of the interval
$I_{IRZ}$ and associated non-Abelian Coulomb phase, in contrast with some of
the values calculated via the conventional loop expansion toward the lower end
of $I_{IRZ}$.  In general, for all of the $N_c$ values considered, our results
for $\gamma_{IR,\Delta_f^p}$ here satisfy the upper bound
(\ref{gamma_upperbound}) and hence are consistent with the conclusion that the
$\Delta_f$ expansion is reasonably reliable throughout the interval $I_{IRZ}$
and non-Abelian Coulomb phase.  We regard this, together with the
scheme-independence itself, as being a major advantage of the $\Delta_f$
expansion.


\subsection{LNN Limit for $\gamma_{\bar\psi\psi,IR}$ }
\label{gamma_lnn_section} 

Here we consider theories with $G={\rm SU}(N_c)$ and $N_f$ copies of 
fermions in the representation $R=F$ in the LNN limit (\ref{lnn}).  We recall
that in this LNN limit, the interval $I_{IRZ}$ is given by
Eq. (\ref{intervalr}) and the scaled $\Delta_r$ is defined by
Eq. (\ref{deltar}).  We define rescaled coefficients $\hat \kappa_n$
\beq
\hat \kappa_n \equiv \lim_{N_c \to \infty} N_c^n \, \kappa_n 
\label{kappahatn}
\eeq
that are finite in this LNN limit. The anomalous dimension
$\gamma_{\bar\psi\psi,IR}$ is also finite in this limit and is given by
\beq
\lim_{LNN} \gamma_{\bar\psi\psi,IR} = \sum_{n=1}^\infty \kappa_n \Delta_f^n 
= \sum_{n=1}^\infty \hat \kappa_n \Delta_r^n \ . 
\label{gamma_ir_lnn}
\eeq
From (\ref{intervalr}), it follows that as $r$ decreases from $r_{b1z}$ to
$r_{b2z}$, $\Delta_r$ increases from 0 to the its maximal value 
\beq
(\Delta_r)_{max} = \frac{75}{26} = 2.8846 \quad {\rm for} \ r \in I_{IRZ,r} \ .
\label{deltarmax}
\eeq

\begin{figure}
  \begin{center}
    \includegraphics[height=6cm]{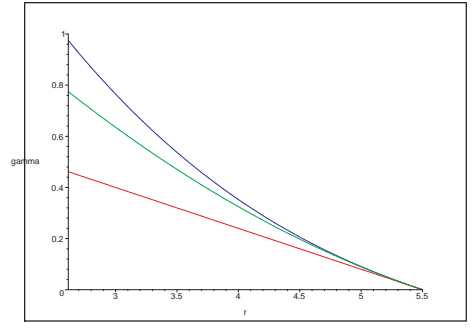}
  \end{center}
\caption{Plot of $\gamma_{\bar\psi\psi,IR,\Delta_r^p}$ for $1 \le p \le 3$ as a
  function of $r \in I_{IRZ,r}$ in the LNN limit (\ref{lnn}).  
 From bottom to top, the curves (with colors online)
 refer to  $\gamma_{\bar\psi\psi,IR,\Delta_r}$ (red),
           $\gamma_{\bar\psi\psi,IR,\Delta_r^2}$ (green)
           $\gamma_{\bar\psi\psi,IR,\Delta_r^3}$ (blue).}
\label{gamma_lnn_plot}
\end{figure}

From the results for $\kappa_n$, $n=1, \ 2, \ 3$ in \cite{gtr} or the special
cases given above for $G={\rm SU}(N_c)$ and $R=F$ in
Eqs. (\ref{kappa1f})-(\ref{kappa3f}), we find
\beq
\hat\kappa_1 = \frac{4}{25} = 0.1600 \ , 
\label{kappahat1}
\eeq
\beq
\hat\kappa_2 = \frac{588}{5^6} = 0.037632 \ , 
\label{kappahat2}
\eeq
and
\beq
\hat\kappa_3 = \frac{2193944}{3^3 \cdot 5^{10}} = 0.83207 \times 10^{-2} \ , 
\label{kappahat3}
\eeq
where, as above, we indicate the factorization of the denominators.
Numerically, to order $O(\Delta_r^3)$, 
\beqs
\lim_{LNN} \gamma_{\bar\psi\psi,IR} & = & 
\Delta_r \Big [ 0.160000 + 0.037632 \Delta_r 
\cr\cr
&+& 0.0083207 \Delta_r^2 + O(\Delta_f^3) \, \Big ] \ . 
\label{gamma_ir_lnn_Delta3}
\eeqs
We plot the value of $\gamma_{\bar\psi\psi,IR}$ calculated to order
$\Delta_r^p$, denoted $\gamma_{\bar\psi\psi,IR,\Delta_r^p}$, 
 for $1 \le p \le 3$, as a function of $r \in I_{IRZ,r}$ in Fig. 
\ref{gamma_lnn_plot}.  As a consequence of the positivity of the
$\hat \kappa_p$ in Eqs. (\ref{kappahat1})-(\ref{kappahat3}), for a fixed $r$, 
$\gamma_{\bar\psi\psi,IR,\Delta_r^p}$ is a monotonically increasing function of
the order of calculation, $p$.  Interestingly, as $r$ decreases toward the
lower end of the interval $I_{IRZ,r}$ at $r=r_{b2z}=34/13 = 2.6154$, the value
of $\gamma_{\bar\psi\psi,IR}$ calculated to the highest order in this LNN
limit, namely $O(\Delta_r^3)$ is slightly less than 1.  This is similar to the
behavior that was found for the specific cases of SU(2) and SU(3) gauge groups
and $R=F$ in \cite{gtr} and for SU(3) with $\gamma_{\bar\psi\psi,IR}$ 
calculated to the next order, $O(\Delta_r^4)$ in \cite{gsi}.  

As discussed above, our calculations of $\gamma_{\bar\psi\psi,IR}$ via the
$\Delta_f$ expansion, both for specific values of $N_c$ and in the LNN limit,
have yielded results satisfying the upper bound (\ref{gamma_upperbound})
throughout the interval $I_{IRZ}$.  These results support the conclusion that
the small-$\Delta_f$ series expansion is reliable throughout this interval
$I_{IRZ}$ and associated non-Abelian Coulomb phase.  It is also worthwhile to
obtain an estimate of the range of applicability of the small-$\Delta_f$ series
expansion via a different method, the aforementioned ratio test.  From the
coefficients $\hat\kappa_n$ that we have calculated with $1 \le n \le 3$, we
compute the ratios
\beq
\frac{\hat\kappa_1}{\hat\kappa_2} = 4.252
\label{kappa1hat_over_kappa2hat}
\eeq
and
\beq
\frac{\hat\kappa_2}{\hat\kappa_3} = 4.523 
\label{kappa2hat_over_kappa3hat}
\eeq
Recalling that the maximal value of $\Delta_r$ in the interval $I_{IRZ,r}$ is
2.885 (Eq. (\ref{Deltar_max_irz}), these ratios are again consistent with the
inference that the small-$\Delta_r$ series expansion may be reasonably accurate
in this interval $I_{IRZ}$. Since $r$ has a maximal value of 5.5 in this LNN
limit, the above ratios also suggest that one could not reliably apply the
small $\Delta_r$ expansion down to small $r$ (see also
\cite{stevenson2016}). This is in agreement with the fact that the properties
of theory change qualitatively as $r$ decreases below $r_c$ in Eq. (\ref{rc});
in particular, there is spontaneous chiral symmetry breaking at small $r$


\subsection{Analysis with Pad\'e Approximants} 
\label{pade_section}

To get further insight into the behavior of $\gamma_{\bar\psi\psi,IR}$, 
we shall calculate
and analyze Pad\'e approximants (PAs) \cite{pades}. For this purpose, we shall 
use the a reduced function normalized to unity at $\Delta_f=0$, namely
\beq
\bar\gamma_{\bar\psi\psi,IR} = \frac{\gamma_{\bar\psi\psi,IR}}
{\kappa_1 \Delta_r} = 
1 + \frac{1}{\kappa_1} \sum_{n=2}^\infty \kappa_n \, \Delta_r^{n-1} \ . 
\label{gammair_reduced}
\eeq
The calculation of $\gamma_{\bar\psi\psi,IR}$ to order 
$\Delta_r^3$ yields $\bar\gamma_{\bar\psi\psi,IR}$
to order $\Delta_r^2$. In turn, from this we can compute three PAs: 
$[2,0]_{\bar\gamma_{\bar\psi\psi,IR}}$, 
$[1,1]_{\bar\gamma_{\bar\psi\psi,IR}}$, and
$[0,2]_{\bar\gamma_{\bar\psi\psi,IR}}$. 
Since the [2,0] PA is just $\bar\gamma_{\bar\psi\psi,IR}$ itself,
to order $\Delta_r^2$, we focus on the [1,1] and [0,2] PAs.  We calculate
\beq
[1,1]_{\bar\gamma_{\bar\psi\psi,IR}} = \frac{1 + \frac{34957}{2480625}\Delta_r}
{1-\frac{548486}{2480625}\Delta_r}
\label{pa11_lnn}
\eeq
and
\beq
[0,2]_{\bar\gamma_{\bar\psi\psi,IR}} = \frac{1}{1 - \frac{147}{625}\Delta_r  
+ \frac{34957}{10546875}\Delta_r^2} \ . 
\label{pa02_lnn}
\eeq

The [1,1] PA has no physical zero and a pole at 
\beq
(\Delta_r)_{pole,[1,1]_{\bar\gamma_{\bar\psi\psi,IR}}} = 
\frac{2480625}{548486} = 4.523
\label{rpole_p11}
\eeq
Since this value is well beyond the maximum value of $\Delta_r$ for
$r \in I_{IRZ,r}$, namely 2.885, it follows that the [1,1] PA is finite for all
$r \in I_{IRZ,r}$.  

The [0,2] PA obviously has no zero, and has two poles, at 
\beqs
(\Delta_r)_{poles,[0,2]_{\bar\gamma_{\bar\psi\psi,IR}}} & = & 
\frac{1875}{69914}(1323 \pm 17\sqrt{4605} \ ) \cr\cr
& = & 4.5425, \ 66.420 
\label{rpole_p02}
\eeqs
The first of these, at $\Delta_r=4.5425$, is well beyond
$(\Delta_r)_{max}=2.885$ so that the [0,2] PA is finite for all $r \in
I_{IRZ,r}$, and the second is also irrelevant, since it corresponds to the
value, $r=72$, far beyond the AF interval, $r \in [0,34/13]$.  The irrelevance
of these poles in the Pad\'e approximants is in agreement with the conclusion
that we have reached from our other methods that the small-$\Delta_f$ expansion
is reasonably reliable throughout the interval $I_{IRZ}$ and related
non-Abelian Coulomb phase. In Table \ref{gamma_ir_values} we list our results
for $\gamma_{\bar\psi\psi,IR,\Delta_r^3}$, 
$[1,1]_{\gamma_{\bar\psi\psi,IR,\Delta_r^3}}$, and
$[0,2]_{\gamma_{\bar\psi\psi,IR,\Delta_r^3}}$, together with 
$\gamma_{\bar\psi\psi,IR,n\ell}$ with $n=2, \ 3, \ 4$ from \cite{lnn} for 
comparison.

We find that if $r$ is in the upper part of the interval $I_{IRZ,r}$, then 
there is excellent agreement between our higher-loop calculations of
$\gamma_{\bar\psi\psi,IR,3\ell,\overline{\rm MS}}$ and 
$\gamma_{\bar\psi\psi,IR,4\ell,\overline{\rm MS}}$ from \cite{bvh} 
and the present calculations of $\gamma_{\bar\psi\psi,IR,\Delta_r^3}$, 
$[1,1]_{\gamma_{\bar\psi\psi,IR,\Delta_r^3}}$, 
and $[0,2]_{\gamma_{\bar\psi\psi,IR,\Delta_r^3}}$.  
As $r$ decreases in this interval
$I_{IRZ,r}$, the values of the anomalous dimension calculated in the various
different ways begin to exhibit small deviations from each other, and, as
expected, these deviations become larger as $r$ descends toward the lower 
end of the interval $I_{IRZ,r}$.  


\section{Scheme-Independent Calculation of Anomalous Dimension 
$\gamma_{T,IR}$ to $O(\Delta_f^3)$ } 
\label{gammat_section}


\subsection{Calculation for General $G$ and $R$}
\label{gammat_general_section}

In this section we present a scheme-independent calculation of 
the anomalous dimension of the (gauge-invariant) bilinear fermion 
antisymmetric rank-2 Dirac tensor operators evaluated at $\alpha_{IR}$. 
The flavor-nonsinglet and flavor-singlet tensor operators of this type 
are
\beq
J_{2,fns} = \bar\psi T_b \ \sigma_{\mu\nu} \psi 
\label{j2fns}
\eeq
and
\beq
J_{2,fs} = \bar\psi \sigma_{\mu\nu} \psi \ , 
\label{j2fs}
\eeq
where, as defined before, $T_b$, $b=1,...,N_f^2-1$, is a generator of 
algebra of SU($N_f$), and 
\beq
\sigma_{\mu\nu} = \frac{i}{2}[\gamma_\mu, \gamma_\nu]
\label{sigmamunu}
\eeq
is the usual antisymmetric rank-2 Dirac tensor.  As was true of the operators
$J_{0,fns}$ and $J_{0,fs}$, the anomalous dimensions of $J_{2,fns}$ and
$J_{2,fs}$ are equal (e.g., \cite{gracey_gammatensor}), so we will denote both
with the single symbol $\gamma_T$ ($T$ for tensor) and the evaluation at
$\alpha_{IR}$ as $\gamma_{T,IR}$.  The usual power series expansion for
$\gamma_T$ in powers of $a$ is
\beq
\gamma_T = \sum_{\ell=1} c_{T,\ell} \, a^\ell \ . 
\label{gammat_aseries}
\eeq
The $c_{T,\ell}$ have been calculated up to $\ell=3$ loop order in
\cite{gracey_gammatensor,gracey_erratum}. We write the scheme-independent
expansion of this anomalous dimension as
\beq
\gamma_{T,IR} = \sum_{n=1}^\infty \kappa_{T,n} \, \Delta_f^n
\label{gammatseries}
\eeq
and denote the truncation of this series at maximal power $n=p$ as 
$\gamma_{T,IR,\Delta_f^p}$. 

For general gauge group $G$ and fermion representation $R$, using the
three-loop results from \cite{gracey_gammatensor,gracey_erratum} together with
the four-loop beta function coefficients $b_\ell$ with $1 \le \ell \le 4$
\cite{b1,b2,b3,b4}, we calculate the following coefficients in the
scheme-independent expansion of $\gamma_{T,IR}$
\beq
\kappa_{T,1} = -\frac{8C_fT_f}{3C_A D} \ , 
\label{kappat1}
\eeq
\beq
\kappa_{T,2} = -\frac{4C_fT_f^2(259C_A^2+428C_AC_f-528C_f^2)}{9C_A^2 D^3} \ , 
\label{kappat2}
\eeq
\begin{widetext}
\beqs
\kappa_{T,3} & = & \frac{4C_f T_f}{3^5 C_A^4 D^5}\bigg [ 
3C_A T_f^2 \bigg \{ 
     C_A^4(-11319+188160\zeta_3) + 
  C_A^3C_f(-337204+64512\zeta_3) + C_A^2C_f^2(83616-890112\zeta_3) \cr\cr
&+& C_AC_f^3(1385472-354816\zeta_3) + C_f^4(-212960+743424\zeta_3) \bigg \}
-512T_f^2D(-5+132\zeta_3)\frac{d_A^{abcd}d_A^{abcd}}{d_A} \cr\cr
&-& 15488C_A^2D(-11+24\zeta_3)\frac{d_R^{abcd}d_R^{abcd}}{d_A}
+11264C_AT_f D(-4+39\zeta_3)\frac{d_R^{abcd}d_A^{abcd}}{d_A} \bigg ]  \ . 
\label{kappat3}
\eeqs
\end{widetext}
We note that 
\beq
\kappa_{T,1}=-\frac{1}{3} \kappa_1 \ . 
\label{kappa1_m2m0rel}
\eeq
%


\subsection{Evalulation for $G={\rm SU}(N_c)$ and $R=F$}
\label{gammat_sun_section}

As we did with the $\kappa_n$ coefficients, we exhibit the reduction
of these general formulas for the gauge group $G={\rm SU}(N_c)$ with $N_f$ 
fermions in the representation $R=F$. In accordance with
Eq. (\ref{kappa1_m2m0rel}), we obtain 
\beq
\kappa_{T,1,{\rm SU}(N_c),F} = -\frac{4(N_c^2-1)}{3N_c(25N_c^2-11)} \ .
\label{kappat1f}
\eeq
Further,
\beq
\kappa_{T,2,{\rm SU}(N_c),F} = -\frac{4(N_c^2-1)(341N_c^4+50N_c^2-132)}
{3^2N_c^2(25N_c^2-11)^3}
\label{kappat2f}
\eeq
and
\begin{widetext}
\beqs
\kappa_{T,3,{\rm SU}(N_c),F} &=& \frac{8(N_c^2-1)}{3^4 N_c^3(25N_c^2-11)^5}
\bigg [ 23057N_c^8-557686N_c^6+1084692N_c^4-354200N_c^2-13310 \cr\cr
&+&192(25N_c^2-11)(163N_c^4-225N_c^2-22)\zeta_3 \bigg ] \ . 
\eeqs
\end{widetext}
The coefficient $\kappa_{T,1,{\rm SU}(N_c),F}$ is manifestly negative for all
$N_c \ge 2$, and this is also true of 
$\kappa_{T,2,{\rm SU}(N_c),F}$, while we find that 
$\kappa_{T,3,{\rm SU}(N_c),F}$ is positive for all $N_c \ge 2$.  


\subsection{LNN Limit for $\gamma_{T,IR}$ }
\label{gammat_lnn_section}

\begin{figure}
  \begin{center}
    \includegraphics[height=6cm]{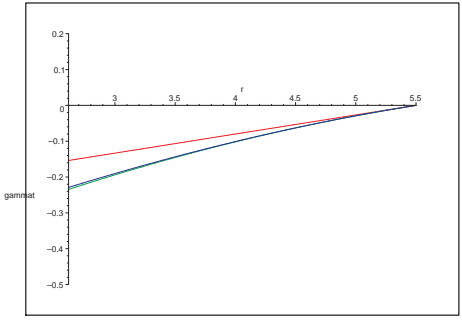}
  \end{center}
\caption{Plot of $\gamma_{T,IR,\Delta_r^p}$ for $1 \le p \le 3$ as a 
  function of $r \in I_{IRZ,r}$ in the LNN limit (\ref{lnn}).  
 From bottom to top, the curves (with colors online)
 refer to  $\gamma_{T,IR,\Delta_r}$ (red),
           $\gamma_{T,IR,\Delta_r^2}$ (green)
           $\gamma_{T,IR,\Delta_r^3}$ (blue).}
\label{gammat_lnn_plot}
\end{figure}

Here we evaluate the $\kappa_{T,n}$ and $\gamma_{T,IR}$ in the LNN limit. 
The rescaled quantities that are finite in this limit are the analogues of
those that we defined and studied for $\gamma_{\bar\psi\psi,IR}$ in 
section \ref{gamma_lnn_section}.  We calculate 
\beq
\hat \kappa_{T,n} = \lim_{N_c \to \infty} N_c^n \kappa_{T,n}
\label{kappat_hat}
\eeq
have the values 
\beq
\hat \kappa_{T,1} = -\frac{4}{3 \cdot 5^2} = -0.053333 \ , 
\label{kappat1hat}
\eeq
\beq
\hat \kappa_{T,2} = -\frac{1364}{3^2 \cdot 5^6} = -(0.969956 \times 10^{-2})
\ , 
\label{kappat2hat}
\eeq
and
\beq
\hat \kappa_{T,3} = \frac{184456}{3^4 \cdot 5^{10}} = 2.3319 \times 10^{-4}
\ . 
\label{kappat3hat}
\eeq
Hence, to third order in the rescaled quantity $\Delta_r$ defined in 
Eq. (\ref{deltar}, we have the following scheme-independent expansion for 
$\gamma_{T,IR}$ in the LNN limit:
\beqs
\lim_{LNN} \gamma_{T,IR} & = & 
\Delta_r \Big [  -0.053333 - (0.96996 \times 10^{-2})\Delta_r 
\cr\cr
&+& (2.3319 \times 10^{-4})\Delta_r^2 + O(\Delta_f^3) \, \Big ] \ . 
\label{gammat_ir_lnn_Delta3}
\eeqs
In Fig. \ref{gammat_lnn_plot} we plot $\gamma_{T,IR,\Delta_r^p}$ for 
$1 \le p \le 3$ as a function of $r$ in the interval $I_{IRZ,r}$.  As a
consequence of the fact that both $\hat\kappa_{T,1}$ and $\hat\kappa_{T,2}$ are
negative, for a fixed value of $r$, $\gamma_{T,IR,\Delta_p^2}$ is negative and
larger in magnitude than $\gamma_{T,IR,\Delta_p^2}$.  Although
$\hat\kappa_{T,3}$ is positive, it is sufficiently small that for a given $r$,
the value of $r$, $\gamma_{T,IR,\Delta_p^3}$ is close to the value of 
$r$, $\gamma_{T,IR,\Delta_p^2}$.  
%


\subsection{Calculation of $\gamma_{T,IR}$ to $O(\Delta_f^3)$ for 
$G={\rm SU}(3)$ and $R=F$ }
\label{gammat_su3_section}

As another interesting comparison, we evaluate our general expressions for the
$\kappa_{T,n}$ in the special case where the gauge group is $G={\rm SU}(3)$ and
the fermion representation is $R=F$.  We find 
\beq
\kappa_{T,{\rm SU}(3),F,1}= 
-\frac{16}{3^2 \cdot 107} = -(1.6615 \times 10^{-2}) \ , 
\label{kappat1_su3f}
\eeq
\beq
\kappa_{T,{\rm SU}(3),F,2} = 
-\frac{37252}{(3 \cdot 107)^3} = -(1.12625 \times 10^{-3}) \ , 
\label{kappat2_su3f}
\eeq
and 
\beqs
\kappa_{T,{\rm SU}(3),F,3} & = & -\frac{341234350}{3^7 \cdot (107)^5} + 
\frac{2855936}{3^6 \cdot (107)^4} \, \zeta_3 \cr\cr
& = & 2.480155 \times 10^{-5} \ . 
\label{kappat3_su3f}
\eeqs
Thus, the leading two terms in the $\Delta_f$ expansion for $J_2$ are negative,
with the coefficient of $\Delta_f^3$ being positive but smaller in magnitude. 
These results may be contrasted to those obtained in \cite{gtr} for 
$\kappa_n \equiv \kappa_{\bar\psi\psi,n}$ with $1 \le n \le 3$ and 
in \cite{gsi} for $n=4$ for this SU(3) theory with $R=F$, which are listed
above in (\ref{kappa1j0su3f})-(\ref{kappa4j0su3f}).  We have computed ratios of
the magnitudes of successive coefficients as before and again infer that the
small-$\Delta_f$ expansion can be reliable in the interval $I_{IRZ}$.


\subsection{Evaluation for $R=adj$}
\label{gammat_adj_section}

For $G={\rm SU}(N_c)$ and $R=adj$, our general results above reduce to 
\beq
\kappa_{T,1,{\rm SU}(N_c),adj}= -\frac{4}{3^3} = -0.05333 \ , 
\label{kappat1_adj}
\eeq
\beq
\kappa_{T,2,{\rm SU}(N_c),adj}= -\frac{53}{2 \cdot 3^7} = 
-(1.2117 \times 10^{-2}) \ , 
\label{kappat2_adj}
\eeq
and
\beq
\kappa_{T,3,{\rm SU}(N_c),adj}= \frac{N_c^2(34799-9216\zeta_3)+42624}
{2^3 \cdot 3^{11}N_c^2} 
\label{kappat3_adj}
\eeq
This is positive for all physical $N_c$ and has the large-$N_c$ limit
\beqs
\lim_{N_c \to \infty} \kappa_{T,3,{\rm SU}(N_c),adj} & = & 
\frac{34799-9216\zeta_3}{2^3 \cdot 3^{11}} \cr\cr
& = & 0.0167381
\label{kappat3_adj_limit}
\eeqs
Thus, the signs of the first three coefficients 
$\kappa_{T,n,{\rm SU}(N_c),adj}$ are the same as those of the 
coefficients $\kappa_{T,n,{\rm SU}(N_c),F}$.  These are summarized in Table
\ref{signs}. 


\section{Conclusions}
\label{conclusion_section}

In conclusion, in this paper we have presented a number of new results on
scheme-independent calculations of various quantities in an asymptotically free
vectorial gauge theory having an IR zero of the beta function.  We consider a
theory with a (non-Abelian) gauge group $G$ and $N_f$ fermions in a
representation $R$ of $G$.  First, we have calculated the derivative
$\beta'_{IR}$, equivalent to $\gamma_{_{F^2,IR}}$, to order $\Delta_f^4$ 
for general $G$ and $R$,
and have given explicit results for $G={\rm SU}(N_f)$ and fermions in the
fundamental and adjoint representations.  For the case $G={\rm SU}(3)$ and
fermions in the fundamental representation, we have also calculated
$\beta'_{IR}$ to the next higher order, $\Delta_f^5$. It would be useful to
have lattice measurements of $\gamma_{_{F^2,IR}}$, which, in the case of SU(3),
could be compared with our calculation of this anomalous dimension.  Second, we
have given more details on the scheme-independent analysis of
$\gamma_{\bar\psi\psi,IR}$ studied earlier in \cite{gtr} and \cite{gsi},
including explicit analytic results for $G={\rm SU}(N_c)$ with fermions in the
fundamental and adjoint representations.  In the former case, we have also
investigated the LNN limit (\ref{lnn}), calculated Pad\'e approximants, and
compared with results from the conventional higher-loop calculation of this
anomalous dimension. Our results are useful for comparisons with lattice
measurements of $\gamma_{\bar\psi\psi,IR}$ and for the fundamental question of
the value of $N_{f,cr}$ and whether $\gamma_{\bar\psi\psi,IR}$ saturates its
upper bound at the lower end of the conformal non-Abelian Coulomb
phase. Moreover, the type of theory considered here may be relevant for
ultraviolet extensions of the Standard Model.  Third, we have presented a
scheme-independent calculation to order $\Delta_f^3$ of the anomalous dimension
$\gamma_{T,IR}$ of the (flavor-nonsinglet and flavor-singlet) bilinear fermion
antisymmetric rank-2 Dirac tensor operators.  We have shown that our
scheme-independent calculations of the anomalous dimensions of 
${\rm Tr}(F_{\mu\nu} F^{\mu\nu})$ and various fermion bilinear operators in the
non-Abelian Coulomb phase obey respective rigorous upper bounds for conformally
invariant theories.  This, together with other inputs including Pad\'e
approximants indicates that the series expansions in powers of $\Delta_f$
should be reasonably accurate throughout the non-Abelian Coulomb phase.  We
believe that the results presented here show the value of scheme-independent
expansions of quantities evaluated at an infrared zero of the beta function in
gauge theories.


This research was supported in part by the Danish National
Research Foundation grant DNRF90 to CP$^3$-Origins at SDU (T.A.R.) and 
by the U.S. NSF Grant NSF-PHY-16-1620628 (R.S.) 


\begin{appendix}


\section{Series Coefficients for $\beta_\xi$ and 
$\gamma_{\bar\psi\psi}$ in the LNN Limit}
\label{bellhatappendix} 

For reference, we list here the rescaled series coefficients for 
$\beta_\xi$ and $\gamma_{\bar\psi\psi}$ in the LNN limit (\ref{lnn}).  
First, we recall that \cite{b1}
\beq
b_1 = \frac{1}{3}(11C_A-4T_fN_f)
\label{b1}
\eeq
and \cite{b2}
\beq
b_2 = \frac{1}{3}\Big [34 C_A^2 - 4(5C_A+3C_f)T_fN_f \Big ] \ , 
\label{b2}
\eeq
where $C_A$, $C_f$, and $T_f$ are group invariants \cite{casimir}. 
It follows that in the LNN limit the $\hat b_\ell$ with $\ell=1,2$ are 
\beq
\hat b_1 = \frac{1}{3}(11-2r)
\label{b1hat}
\eeq
and
\beq
\hat b_2 = \frac{1}{3}(34-13r)  \ .
\label{b2hat}
\eeq
The coefficients $b_3$ and $b_4$ have been calculated in the 
$\overline{\rm MS}$ scheme \cite{b3,b4}. 
With these inputs, one obtains \cite{lnn}
\beq
\hat b_3 = \frac{1}{54}(2857-1709r+112r^2)
\label{b3hat}
\eeq
and 
\beqs
\hat b_4 & = & \frac{150473}{486}-  \Big ( \frac{485513}{1944} \Big ) r 
+ \Big ( \frac{8654}{243} \Big ) r^2 \cr\cr
&+&\Big ( \frac{130}{243}  \Big ) r^3 + \frac{4}{9}(11-5r+21r^2) \,  \zeta_3
\ . 
\label{b4hat}
\eeqs

For the coefficients $\hat c_\ell$ in Eq. (\ref{gamma_ir_lnn}), one has 
(\cite{c4} and references therein) 
\beq
\hat c_1 = 3 \ , 
\label{chat1}
\eeq
\beq
\hat c_2 = \frac{203}{12} - \frac{5}{3} r \ , 
\label{chat2}
\eeq
\beq
\hat c_3 = \frac{11413}{108}-\bigg ( \frac{1177}{54} + 12\zeta_3 \bigg ) r
- \frac{35}{27}r^2 \ , 
\label{chat3}
\eeq
and
\beqs
\hat c_4 & = & \frac{460151}{576}-\frac{23816}{81}r+\frac{899}{162}r^2
-\frac{83}{81}r^3 \cr\cr
&+& \bigg (\frac{1157}{9}-\frac{889}{3}r+20r^2+\frac{16}{9}r^3 \bigg )\zeta_3
\cr\cr 
&+& r\Big (66-12r \Big )\zeta_4 + \Big (-220+160r \Big )\zeta_5 \ . 
\label{chat4}
\eeqs

\end{appendix}



\newpage

\begin{table}
\caption{\footnotesize{Signs of expansion coefficients discussed in the text
for gauge group $G={\rm SU}(N_c)$ and fermion representation $R=F$
(fundamental) and $R=adj$ (adjoint). 
Several results on signs actually apply more generally for arbitrary $G$ and 
$R$; see text for details. For $G={\rm SU}(3)$, 
we have also calculated $d_{5,F}$ in Eq. (\ref{d5su3}) and find that it is 
negative. The entry for $\kappa_{4,F}$ applies for $G={\rm SU}(3)$ 
(see Eq. (\ref{kappa4j0su3f})), as calculated in \cite{gsi}, 
and this is indicated by the $(*)$. 
The entry NA means ``not available'', i.e. the coefficient has not yet been 
calculated.}}
\begin{center}
\begin{tabular}{|c|c|c|c|c|c|c|} \hline\hline 
$n$ & $d_{n,F}$ & $d_{n,adj}$ & $\kappa_{n,F}$ & $\kappa_{n,adj}$ 
& $\kappa_{_{T,n,F}}$ & $\kappa_{_{T,n,adj}}$ 
\\ \hline
 1   & 0    &  0    &  $+$    & $+$  & $-$  & $-$  \\
 2   & $+$  &  $+$  &  $+$    & $+$  & $-$  & $-$  \\
 3   & $+$  &  $+$  &  $+$    & $+$  & $+$  & $+$  \\
 4   & $-$  &  $+$  &  $+$(*)  & NA   & NA   & NA   \\
\hline\hline
\end{tabular}
\end{center}
\label{signs}
\end{table}


\begin{table}
\caption{\footnotesize{ Scheme-independent values of
$\beta'_{IR,\Delta_f^p}$ with $2 \le p \le 4$ for $G={\rm SU}(2)$, SU(3), 
and SU(4), as functions of $N_f$ in the respective intervals $I_{IRZ}$ given in
Eq. (\ref{nfinterval}) with (\ref{nfb1z}) and (\ref{nfb2z}). For 
comparison, we list the $n$-loop values of $\beta'_{IR,n\ell}$ with 
$2 \le n \le 4$, where $\beta'_{IR,3\ell}$ and $\beta'_{IR,4\ell}$ are computed
in the $\overline{\rm MS}$ scheme. Values that exceed the upper bound 
(\ref{gammaff_upperbound}) are marked as such. In the case
of SU(3), we also include our calculation of $\beta'_{IR,\Delta_f^5}$. 
The notation $a$e-$n$ means $a \times 10^{-n}$.
The notation $-$ means that the entry has not been calculated.}}
\begin{center}
\begin{tabular}{|c|c|c|c|c|c|c|c|c|} \hline\hline
$N_c$ & $N_f$ 
    & $\beta'_{IR,2\ell}$
    & $\beta'_{IR,3\ell,{\overline{\rm MS}}}$
    & $\beta'_{IR,4\ell,{\overline{\rm MS}}}$
    & $\beta'_{IR,\Delta_f^2}$
    & $\beta'_{IR,\Delta_f^3}$
    & $\beta'_{IR,\Delta_f^4}$
    & $\beta'_{IR,\Delta_f^5}$ 
\\ \hline
 2 &  6 & $>3$   & 1.620 & 0.975 & 0.499   & 0.957   & 0.734  & $-$ \\
 2 &  7 & 1.202  & 0.728 & 0.677 & 0.320   & 0.554   & 0.463  & $-$ \\
 2 &  8 & 0.400  & 0.318 & 0.300 & 0.180   & 0.279   & 0.250  & $-$ \\
 2 &  9 & 0.126  & 0.115 & 0.110 & 0.0799  & 0.109   & 0.1035 & $-$ \\
 2 & 10 & 0.0245 & 0.0239& 0.0235& 0.0200  & 0.0236  & 0.0233 & $-$ \\
 \hline
 3 &  9 & $>3$   & 1.475  & 1.464  & 0.467   & 0.882   & 0.7355  & 0.602 \\
 3 & 10 & 1.523  & 0.872  & 0.853  & 0.351   & 0.621   & 0.538   & 0.473 \\
 3 & 11 & 0.720  & 0.517  & 0.498  & 0.251   & 0.415   & 0.3725  & 0.344 \\
 3 & 12 & 0.360  & 0.2955 & 0.282  & 0.168   & 0.258   & 0.239   & 0.228 \\
 3 & 13 & 0.174  & 0.1556 & 0.149  & 0.102   & 0.144   & 0.137   & 0.134 \\
 3 & 14 & 0.0737 & 0.0699 & 0.678  & 0.0519  & 0.0673  & 0.0655  & 0.0649 \\
 3 & 15 & 0.0227 & 0.0223 & 0.0220 & 0.0187  & 0.0220  & 0.0218  & 0.0217 \\
 3 & 16 & 2.21e-3& 2.20e-3& 2.20e-3& 2.08e-3 & 2.20e-3 & 2.20e-3& 2.20e-3 \\
\hline
 4 & 11 & $>3$   & 2.189 & 2.189  & 0.553   & 1.087   & 0.898  & $-$ \\
 4 & 12 & $>3$   & 1.430 & 1.429  & 0.457   & 0.858   & 0.729  & $-$ \\
 4 & 13 & 1.767  & 0.965 & 0.955  & 0.370   & 0.663   & 0.578  & $-$ \\
 4 & 14 & 0.984  & 0.655 & 0.639  & 0.292   & 0.498   & 0.445  & $-$ \\
 4 & 15 & 0.581  & 0.440 & 0.424  & 0.224   & 0.362   & 0.331  & $-$ \\
 4 & 16 & 0.348  & 0.288 & 0.276  & 0.1645  & 0.251   & 0.234  & $-$ \\
 4 & 17 & 0.204  & 0.180 & 0.1725 & 0.114   & 0.164   & 0.156  & $-$ \\
 4 & 18 & 0.113  & 0.105 & 0.101  & 0.0731  & 0.0988  & 0.0955 & $-$ \\
 4 & 19 & 0.0558 & 0.0536& 0.0522 & 0.0411  & 0.0520  & 0.0509 & $-$ \\
 4 & 20 & 0.0222 & 0.0218& 0.0215 & 0.0183  & 0.0215  & 0.0213 & $-$ \\
 4 & 21 & 5.01e-3&4.99e-3&4.96e-3 & 4.57e-3 & 4.97e-3 & 4.96e-3& $-$ \\
\hline\hline
\end{tabular}
\end{center}
\label{betaprime_values}
\end{table}


\begin{table}
\caption{\footnotesize{ Scheme-independent values of
$\beta'_{IR,\Delta_r^p}$ for $2 \le p \le 4$
in the LNN limit (\ref{lnn}) as functions of $r=5.5-\Delta_r$. For comparison,
we also list the $n$-loop values $\beta'_{IR,n\ell}$ with $2 \le n \le 4$,
where $\beta'_{IR,3\ell}$ and $\beta'_{IR,4\ell}$ are computed in the 
$\overline{\rm MS}$ scheme (and values that exceed the 
upper bound (\ref{gammaff_upperbound}) are marked as such). 
 The notation $a$e-$n$ means $a \times 10^{-n}$. }}
\begin{center}
\begin{tabular}{|c|c|c|c|c|c|c|} \hline\hline
$r$ & $\beta'_{IR,2\ell}$
    & $\beta'_{IR,3\ell,{\overline{\rm MS}}}$
    & $\beta'_{IR,4\ell,{\overline{\rm MS}}}$
    & $\beta'_{IR,\Delta_r^2}$
    & $\beta'_{IR,\Delta_r^3}$
    & $\beta'_{IR,\Delta_r^4}$
\\ \hline
 2.8  & $>3$     & 1.918    & 1.949   & 0.518     & 1.004     & 0.851  \\
 3.0  & $>3$     & 1.376    & 1.523   & 0.444     & 0.830     & 0.717  \\
 3.2  & 1.856    & 1.006    & 1.100   & 0.376     & 0.676     & 0.596  \\
 3.4  & 1.153    & 0.7395   & 0.72985 & 0.314     & 0.542     & 0.486  \\
 3.6  & 0.752    & 0.542    & 0.528   & 0.257     & 0.426     & 0.388  \\
 3.8  & 0.500    & 0.393    & 0.378   & 0.2055    & 0.327     & 0.303  \\
 4.0  & 0.333    & 0.279    & 0.267   & 0.160     & 0.243     & 0.229  \\
 4.2  & 0.219    & 0.193    & 0.185   & 0.120     & 0.174     & 0.166  \\
 4.4  & 0.139    & 0.128    & 0.122   & 0.0860    & 0.119     & 0.115  \\
 4.6  & 0.0837   & 0.0792   & 0.0766  & 0.0576    & 0.0756    & 0.0737 \\
 4.8  & 0.0460   & 0.0445   & 0.0435  & 0.0348    & 0.0433    & 0.0426 \\
 5.0  & 0.0215   & 0.0212   & 0.0208  & 0.0178    & 0.0209    & 0.0207 \\
 5.2  & 0.714e-2 & 0.710e-2 & 0.706e-2& 0.640e-2  & 0.707e-2  & 0.704e-2 \\
 5.4  & 0.737e-3 & 0.736e-3 & 0.7356e-3&0.7111e-3  & 0.7358e-3& 0.7355-3 \\
\hline\hline
\end{tabular}
\end{center}
\label{betaprime_values_lnn}
\end{table}


%
\begin{table}
\caption{\footnotesize{Values of the anomalous dimension
 $\gamma_{\bar\psi\psi,IR,\Delta_f^p}$
    calculated to order $p=1, \ 2, \ 3$, for $G={\rm SU}(N_c)$ and 
$R=F$, as functions of $N_c$ and $N_f$.  To save space, we omit the
subscript $\bar\psi\psi$, writing 
$\gamma_{\bar\psi\psi,IR,\Delta_f^p} \equiv \gamma_{IR,\Delta_f^p}$. 
For comparison, we also include the (scheme-independent) 
$\gamma_{IR,2\ell}$ and 
 $\gamma_{IR,n\ell,\overline{\rm MS}}$, $n=3, \ 4$. 
 $\gamma_{IR,4\ell,\overline{\rm MS}}$.  Values that exceed the 
 bound $\gamma_{\bar\psi\psi,IR} < 2$ in Eq. (\ref{gamma_upperbound})) 
are marked as such; in these cases, 
the $\gamma_{IR,n\ell,\overline{\rm MS}}$ are unlisted (u).}}
\begin{center}
\begin{tabular}{|c|c|c|c|c|c|c|c|} \hline\hline
$N_c$ & $N_f$ & $\gamma_{IR,2\ell}$ & $\gamma_{IR,3\ell,\overline{\rm MS}}$ & 
$\gamma_{IR,4\ell,\overline{\rm MS}}$ & $\gamma_{IR,\Delta_f}$ & 
$\gamma_{IR,\Delta_f^2}$ & $\gamma_{IR,\Delta_f^3}$  
\\ \hline
 2  &  6  & $>2$   & u      & u      & 0.337  & 0.520   & 0.596   \\
 2  &  7  & $>2$   & u      & u      & 0.270  & 0.387   & 0.426   \\
 2  &  8  & 0.752  & 0.272  & 0.204  & 0.202  & 0.268   & 0.285   \\
 2  &  9  & 0.275  & 0.161  & 0.157  & 0.135  & 0.164   & 0.169   \\
 2  & 10  & 0.0910 & 0.0738 & 0.0748 & 0.0674 & 0.07475 & 0.07535 \\
 \hline
 3  &  9  & $>2$   & u      & u      & 0.374  & 0.587  & 0.687   \\
 3  & 10  & $>2$   & u      & u      & 0.324  & 0.484  & 0.549   \\
 3  & 11  & 1.61   & 0.439  & 0.250  & 0.274  & 0.389  & 0.428   \\
 3  & 12  & 0.773  & 0.312  & 0.253  & 0.224  & 0.301  & 0.323   \\
 3  & 13  & 0.404  & 0.220  & 0.210  & 0.174  & 0.221  & 0.231   \\
 3  & 14  & 0.212  & 0.146  & 0.147  & 0.125  & 0.148  & 0.152   \\
 3  & 15  & 0.0997 & 0.0826 & 0.0836 & 0.0748 & 0.0833 & 0.0841  \\
 3  & 16  & 0.0272 & 0.0258 & 0.0259 & 0.0249 & 0.0259 & 0.0259  \\
\hline
 4  & 11  & $>2$   & u      & u      & 0.424  & 0.694  & 0.844   \\
 4  & 12  & $>2$   & u      & u      & 0.386  & 0.609  & 0.721   \\
 4  & 13  & $>2$   & u      & u      & 0.347  & 0.528  & 0.610   \\
 4  & 14  & $>2$   & u      & u      & 0.308  & 0.451  & 0.509   \\
 4  & 15  & 1.32   & 0.420  & 0.281  & 0.270  & 0.379  & 0.418   \\
 4  & 16  & 0.778  & 0.325  & 0.269  & 0.231  & 0.312  & 0.336   \\
 4  & 17  & 0.481  & 0.251  & 0.234  & 0.193  & 0.249  & 0.263   \\
 4  & 18  & 0.301  & 0.189  & 0.187  & 0.154  & 0.190  & 0.197   \\
 4  & 19  & 0.183  & 0.134  & 0.136  & 0.116  & 0.136  & 0.139   \\
 4  & 20  & 0.102  & 0.0854 & 0.0865 & 0.0771 & 0.0860 & 0.086   \\
 4  & 21  & 0.0440 & 0.0407 & 0.0409 & 0.0386 & 0.0408 & 0.0409  \\
\hline\hline
\end{tabular}
\end{center}
\label{gamma_values}
\end{table}



\begin{table}
\caption{\footnotesize{Values of 
$\gamma_{\bar\psi\psi,IR,\Delta_r^3}$, 
$[1,1]_{\gamma_{\bar\psi\psi,IR,\Delta_r^3}}$, and
 $[1,1]_{\gamma_{\bar\psi\psi,IR,\Delta_r^3}}$, together with 
$\gamma_{\bar\psi\psi,IR,n\ell}$ with $n=2, \ 3, \ 4$ from Table V of 
\cite{lnn} for comparison, as a
function of $r$ for $r \in I_{IRZ,r}$ and satisfying $\gamma_{IR} < 2$.
Here, $\Delta_r=5.5-r$, as in Eq. (\ref{deltar}). 
To save space, we omit the subscript $\bar\psi\psi$ below. 
Values that exceed the bound $\gamma_{\bar\psi\psi,IR} < 2$ from conformal 
invariance (see Eq. (\ref{gamma_upperbound})) are marked as such.}}
\begin{center}
\begin{tabular}{|c|c|c|c|c|c|c|} \hline\hline
$r$ & $\gamma_{_{IR,2\ell}}$ & $\gamma_{_{IR,3\ell}}$ & $\gamma_{_{IR,4\ell}}$
& $\gamma_{IR,\Delta_r^3}$ 
& $[1,1]_{\gamma_{IR,\Delta_r^3}}$ 
& $[0,2]_{\gamma_{IR,\Delta_r^3}}$ 
\\ \hline
 2.8   &  $>2$     &  1.708      &  0.1902  & 0.8701  & 1.1127  & 1.1102  \\
 3.0   &  $>2$     &  1.165      &  0.2254  & 0.7652  & 0.9259  & 0.9244  \\
 3.2   &  $>2$     &  0.8540     &  0.2637  & 0.6683  & 0.7731  & 0.7722  \\
 3.4   &  $>2$     &  0.6563     &  0.2933  & 0.5790  & 0.6458  & 0.6453  \\
 3.6   &  1.853    &  0.5201     &  0.3083  & 0.4969  & 0.5383  & 0.5380  \\
 3.8   &  1.178    &  0.4197     &  0.3061  & 0.4216  & 0.4463  & 0.4461  \\
 4.0   &  0.7847   &  0.3414     &  0.2877  & 0.3528  & 0.3667  & 0.3666  \\
 4.2   &  0.5366   &  0.2771     &  0.2566  & 0.2899  & 0.2973  & 0.2972  \\
 4.4   &  0.3707   &  0.2221     &  0.2173  & 0.2326  & 0.2362  & 0.23615 \\
 4.6   &  0.2543   &  0.1735     &  0.1745  & 0.1805  & 0.18205 & 0.18205 \\
 4.8   &  0.1696   &  0.1294     &  0.1313  & 0.1333  & 0.1338  & 0.1338  \\
 5.0   &  0.1057   &  0.08886    &  0.08999 & 0.09045 & 0.09058 & 0.09058 \\
 5.2   &  0.05620  &  0.05123    &  0.05156 & 0.05161 & 0.05163 & 0.05163 \\
 5.4   &  0.01682  &  0.01637    &  0.01638 & 0.01638 & 0.01638 & 0.01638 \\
\hline\hline
\end{tabular}
\end{center}
\label{gamma_ir_values}
\end{table}



\begin{thebibliography}{99}
\bibitem{nfintegral}
%
Here and below, if an expression for $N_f$
formally evaluates to a non-integral real value, it is understood
implicitly that one infers an appropriate integral value from it.
\bibitem{gammaconvention}
Some authors use the opposite sign convention for the anomalous dimension, 
writing $D_{\cal O} = D_{\cal O,{\rm free}} + \gamma_{\cal O}$. Our sign 
convention is the same as the one used in \cite{wtc} and the lattice gauge
theory literature.
\bibitem{higherderivs}
%
The higher derivatives of the beta function evaluated at $\alpha_{IR}$, 
$d^k \beta/(d\alpha)^k|_{\alpha=\alpha_{IR}}$ with $k \ge 2$ are, in general,
scheme-dependent.
\bibitem{otherform}
%
In terms of the rescaled $F^a_{\mu\nu,r} \equiv gF^a_{\mu\nu}$,
Eq. (\ref{traceanomaly}) is expressed equivalently as
$T^\mu_\mu = [\beta/(16\pi \alpha^2)]F^a_{\mu\nu,r}F^{a \ \mu\nu}_r$.
\bibitem{glueball}
%
The operator $F^2 \equiv F_{\mu\nu}^a F^{a \ \mu\nu}$ is the glueball operator
in the phase with confinement and color-singlet hadrons and glueballs, but
there are no glueballs or hadrons in the non-Abelian Coulomb phase, which is
deconfined.  (The context will always make clear when the symbol $F$ refers to
the fundamental representation of $G$ and when it is used in the abbreviation
$F^2$ above.)
\end{thebibliography}
\end{document}